# Fully Integrated Memristive Spiking Neural Network with Analog Neurons for High-Speed Event-Based Data Processing

Zhu Wang[1,3,#], Song Wang[1,#], Zhiyuan Du[1,3], Ruibin Mao[1,3], Yu Xiao[2], Hayden Kwok-Hay So[1], Peng Lin[2]*, and Can Li[1,3]*

[1]Department of Electrical and Electronic Engineering, The University of Hong Kong, Pokfulam, Hong Kong SAR, China

[2]College of Computer Science and Technology, Zhejiang University, Hangzhou, China

[3]Centre for Advanced Semiconductors and Integrated Circuits, The University of Hong Kong, Pokfulam, Hong Kong SAR, China

[#]These authors contributed equally

Email: penglin@zju.edu.cn; canl@hku.hk

## Abstract

The demand for edge artificial intelligence to process event-based, complex data calls for hardware beyond conventional digital, von-Neumann architectures. Neuromorphic computing, using spiking neural networks (SNNs) with emerging memristors, is a promising solution, but existing systems often discard temporal information, demonstrate non-competitive accuracy, or rely on neuron designs with large capacitors that limit the scalability and processing speed. Here we experimentally demonstrate a fully integrated memristive SNN with a 128×24 memristor array integrated on a CMOS chip and custom-designed analog neurons, achieving high-speed, energy-efficient event-driven processing of accelerated spatiotemporal spike signals with high computational fidelity. This is achieved through a proportional time-scaling property of the analog neurons, which allows them to use only compact on-chip capacitors and train directly on the spatiotemporal data without special encoding by backpropagation through surrogate gradient, thus overcoming the speed, scalability and accuracy limitations of previous designs. We experimentally validated our hardware using the DVS128 Gesture dataset, accelerating each sample 50,000-fold to a 30 μs duration. The system achieves an experimental accuracy of 93.06% with a measured energy efficiency of 101.05 TSOPS/W. We project significant future efficiency gains by leveraging picosecond-width spikes and advanced fabrication nodes. By decoupling the hardware's operational timescale from the data's natural timescale, this work establishes a viable pathway for developing neuromorphic processors capable of high-throughput analysis, critical for rapid-response edge computing applications like high-speed analysis of buffered sensor data or ultra-fast in-sensor machine vision.

# Introduction

The escalating demands for real-time artificial intelligence at the edge, processing complex, dynamic, and sometimes sparse, spiking data from event-based sensors [1-3] under strict power and latency constraints, call for hardware beyond conventional digital von Neumann architectures [4-6]. Neuromorphic computing, particularly spiking neural networks (SNN) implemented with emerging non-volatile resistive memories (RRAM), aka., memristors, offers a promising brain-inspired alternative, showing orders of magnitude improvements in energy efficiency through in-memory, event-driven analog computation [7, 8]. There has been intensive research on memristor-based analog in-memory computing [9-12], but most existing reports rely on level-based analog signal processing with artificial neural network (ANN), which are not inherently event-driven. This requires a certain duration of sustained voltage input pulses to stabilize the output current for reliable readout by transimpedance amplifiers and costly analog-to-digital conversion. Accordingly, despite already showing energy and latency improvements over conventional approaches, this incurs a major portion of overhead and fails to fully exploit the intrinsic speed advantages of memristor crossbars and the potential efficiency of sparse, event-based processing.

In response, there have been intensive research efforts in implementing SNNs in memristor-based hardware [13-17], but a fully integrated implementation that achieves practical, high-accuracy inference for general applications has been lacking. Pioneer research has explored using emerging devices to emulate elements in biological neural networks, including synapse dynamics [13, 18-20], such as spike-timing-dependent plasticity (STDP) for unsupervised learning [21], and neuron dynamics [22-25], such as leaky-integrate-and fire [26]; While valuable for fundamental understanding, these systems are typically tailored for specialized scenarios and have not yet achieved the robust, general-purpose accuracy for broad applications, due to challenges on both algorithm and emerging device performance such as limited cycling endurance and high energy and speed overhead for state update. Consequently deployment at the current stage needs to focus on deploying an offline trained model to memristor-based synapses, where the devices do not need to be frequently updated, while using CMOS technology for analog spiking neuron circuits. Yet, even with this pragmatic approach, leveraging CMOS neurons, existing memristive SNN system proposals encounter significant bottlenecks. Many implementations use rate-based coding schemes, which discard the temporally dependent information, thereby degrading accuracy performance and incurring latency [27]. In addition, previous analog SNN systems usually required their neuronal time constants to be aligned with the specific, often slow, natural time scale of the task they address (typically in the millisecond range) [28, 29]. Implementing such long constants with CMOS-based analog neurons will require either large on-chip capacitors (hence a significant chip area for a scaled system) or ultra-low sub-threshold bias currents [30]. However, operating in these time constants prevents these SNN systems from high-speed processing recorded, accelerated data not directly from sensors, or processing general-purpose data converted into spiking form, heavily limiting their application scenarios. Accordingly, a physical experiment on a fully integrated memristive SNN system that achieves high-speed inference with state-of-the-art accuracy has not been reported.

Here, we introduce and experimentally implement a fully integrated memristive SNN system that overcome these bottlenecks. Our system includes a 128×24 memristor crossbar synapse network integrated with customized-designed analog CMOS spike response model (SRM) neurons, designed and fabricated using 180-nm technology node. Our neuron design employs physical computation through resistor-capacitor filters to implement the SRM kernels and is custom-designed to allow fast inference of accelerated input event streams with small capacitors by leveraging the proportional time-scaling property. All kernels are designed to operate after the synaptic network, ensuring that only short pulses pass through memristor devices, thus saving energy owing to the zero static energy consumption of memristor devices. Operating solely on asynchronous input events, our system performs direct temporal stream processing that fully exploits the temporal dynamics within input streams without relying on rate-based artificial neural coding schemes to improve accuracy, thereby reducing the associated coding latency. The SNN is trained with surrogate gradient descent optimized for our fabricated SRM neurons, and we showcased the high-speed inference and direct temporal stream processing with our integrated SNN system. To validate our approach, we demonstrate high-speed spatiotemporal pattern recognition on the DVS128 Gesture dataset, with input signals accelerated 50,000-fold, for processing a gesture sample within 44.31 μs. Our fabricated chip achieves an experimental classification accuracy of 93.06% and an energy efficiency of 101.05 TSOPS/W, with a clear pathway to significant further gains by leveraging picosecond-width spikes and advanced fabrication nodes. This work represents the first physical demonstration of a fully integrated, analog memristive SNN achieving state-of-the-art accuracy on a complex, temporally encoded task, establishing a viable path towards ultra-fast and efficient machine intelligence at the edge.

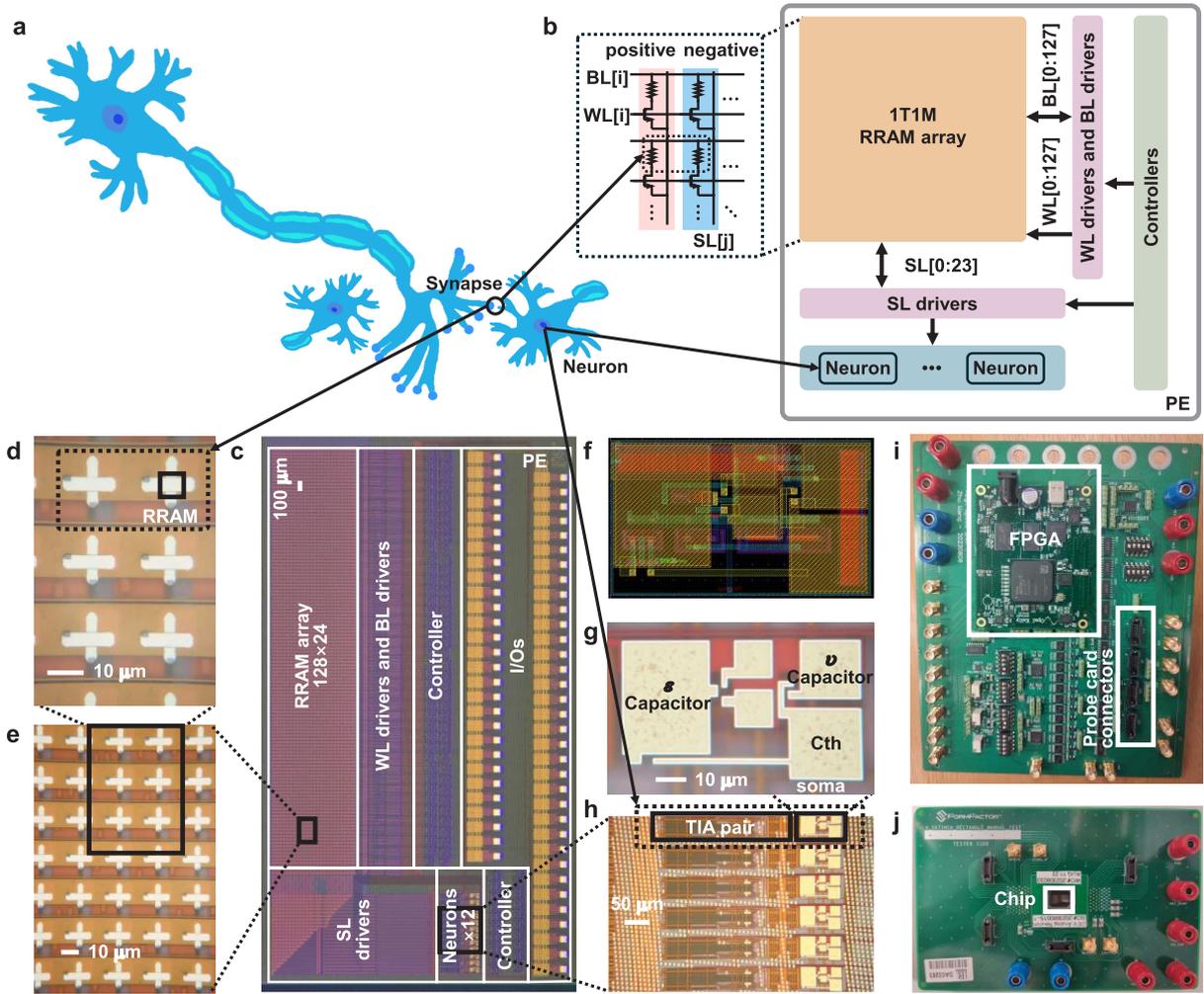

**Fig. 1 | Overview of fully-analog spiking neural networks with integrated memristor synapses and CMOS neurons.** (a) Representation of biological neural networks comprising synapses and neurons. (b) Overview chip architecture for a processing element (PE) macro integrated with a 1T1M crossbar using a differential-pair encoding scheme (inset) as synapses, CMOS neurons, drivers, and controllers. (c) Optical micrograph of the fabricated macro chip using a 180-nm technology node. (d, e) Zoomed-in views of the 128×24 RRAM array and an individual memristor device (indicated by a solid square in *d*). (f) Layout of the soma circuit of an SRM neuron. (g) Micrograph of the neuron's soma circuit showing the capacitors for the $\varepsilon$ and $\upsilon$ kernels and a Cth capacitor. (h) Details of the neuron array, with each neuron (indicated by a dotted line) containing a transimpedance amplifier (TIA) pair and a soma circuit. (i, j) Measurement board (i) that controls and communicates with the chip via a probe card (j), which interfaces the measurement board with the chip.

## SNN system architecture

Our SNN system resembles the structure in biological neural networks, where neurons are interconnected by synapses (Fig. 1a). The neuron circuit is implemented with CMOS analog circuit, while the synaptic network is implemented with one-transistor one-memristor (1T1M) arrays. The CMOS circuits, including neurons, drivers, controllers and the access transistors within the 1T1M array, are designed and fabricated in a commercial foundry using a 180 nm technology node. Ta/TaOx/Pt memristors are then integrated on top of the access transistors using in-house back-end processes in our university cleanroom. Fig. 1b shows the architecture design of our demo chip. During inference, pre-synaptic spikes, in the form of short input voltage pulses, are delivered to the horizontal bit-lines (BLs) of the 1T1M crossbar. All transistors in the crossbar are turned ON, and vertical source-lines (SLs) are virtually grounded by neurons, and consequently, the voltage pulses are weighted based on the conductance of memristors, according to circuit laws. This results in post-synaptic spikes as short pulses with varying amplitude. These resultant spikes are then fed into the analog neuron circuits, which in turn generate new output spikes. These outputs can either represent the final inferred results or be propagated as inputs to subsequent layers in a deeper network. Noteworthy, only short pulses are applied to the memristor devices, thereby minimizing energy consumption, which resembles biological systems. Secondly, because all information is processed entirely within the analog domain, our system bypasses the need for power-intensive and area-consuming analog-to-digital converters.

Fig. 1c presents an optical micrograph of the fabricated demo chip, with a detailed view of a portion of the memristor synapse array in Fig. 1d and Fig. 1e. The layout design of our fabricated neuron circuits is shown in Fig. 1f, while Fig. 1g presents corresponding microscope images, and Fig. 1h shows a zoomed out image that includes a TIA pair that converts the difference of current spikes to voltage spikes. (Some details in these microscope images are obscured by non-transparent metal layers but are visible in the layout design of Fig. 1f). The design, wherein a pair of SL drivers and a following neuron fit within the width of paired columns of the array, and word-line (WL) and BL drivers fit within the height of each row, enables the scalability of the RRAM array to larger dimensions.

Programming and measurement are conducted using a measurement board (Fig. 1i), which is interfaced with the chip through a probe card (Fig. 1j). An onboard field-programmable gate array (FPGA) controls the RRAM programming and manages communication with the chip. We use an arbitrary waveform generator (AWG) to generate high-speed event streams, which are delivered to the chip. Output spike streams from the neurons are captured by the FPGA and transmitted to a computer for subsequent decision-making in the software domain. More details about the chip design and the board design can be found in Methods.

## Custom-designed SRM neurons

Spiking neuron design is central to our SNN system, because of the need to process information encoded in the precise timing of spikes. The biophysically detailed Hodgkin-Huxley (HH) neuron model [31] accurately replicates realistic neuronal behaviour but incurs significant computational complexity, and its effectiveness in applications remains unclear. In contrast, while the simpler Leaky Integrate-and-Fire (LIF) model is more efficient [26], its limited dynamics can constrain performance in tasks requiring complex temporal pattern recognition [32-35]. The spike response model (SRM) [36, 37] can emulate richer dynamics and evidence suggests these can improve performance in tasks where temporal information is critical. However, SRMs are significantly more computationally expensive than LIF when implemented in digital hardware, which has limited their popularity.

In this work, we implement the SRM neurons using analog circuits to address these limitations. Our approach leverages simple passive resistor-capacitor (RC) filters to realize the temporal convolution kernels, which are computationally expensive in digital hardware. We further custom-designed the SRM neuron for more efficient processing particularly with the memristor-based hardware: First, the kernel was engineered so that, when implemented using analog filters, its time constants and the temporal scale of input spike streams maintain proportionality by the same factor, thereby preserving its fundamental response characteristics. This allows the system to process event data accelerated to microsecond-scale durations using compact on-chip capacitors (e.g., 2 pF and 700 fF for time constants), overcoming a critical bottleneck in analog neuron scalability where large capacitances are typically needed for slower, biologically relevant time constants. Second, we integrated the synapse filter as part of the neuron design. While mathematically equivalent to a separate pre-synapse filter, this integration provides a crucial benefit when implemented in memristor hardware: only short voltage pulses (spikes), instead of convolved dynamic voltages, are applied to the memristors. This significantly reduces energy consumption.

Mathematically, the forward propagation of the SNN with our SRM neuron within layer $l$ is expressed as

$$\boldsymbol{o}^{(l)}(t) = \boldsymbol{s}^{(l)}(t)\boldsymbol{W}^{(l)}, \quad (1)$$

$$\boldsymbol{u}^{(l)}(t) = \int_0^t \varepsilon(t-\tau)\boldsymbol{o}^{(l)}(\tau)d\tau, \quad (2)$$

$$\boldsymbol{s}^{(l+1)}(t) = f_s\left(\boldsymbol{u}^{(l)}(t)\right). \quad (3)$$

$$f_s(u): s(t_i) = \begin{cases} \delta(t-t_i), & \text{if } u(t_i) \geq c + \int_0^{t_i} v(t_i-\tau)s^{(l+1)}(\tau)d\tau \\ 0, & \text{otherwise} \end{cases}. \quad (4)$$

Input spikes $\boldsymbol{s}^{(l)}(t) = \sum \delta(t-t_i)$ from previous layer or external source are weighted by memristor synapses $\boldsymbol{W}^{(l)}$ to produce post-synaptic spikes $\boldsymbol{o}^{(l)}(t)$ (Eq. 1). The membrane potential $\boldsymbol{u}^{(l)}(t)$ results from convolving an exponential response kernel $\varepsilon = \frac{1}{\tau_s} e^{-t/\tau_s} \Theta(t)$ (Eq. 2), where $\Theta(t)$ represents the Heaviside step function and $\tau_s$ denotes neuron time constant for the response kernel. Output spikes $s^{(l+1)}(t)$ are generated when $u^{(l)}(t)$ surpasses an adaptive threshold (Eq. 3). The threshold dynamically adjusts via convolution of output spikes with an exponential refractory kernel $\upsilon = \frac{1}{\tau_r} e^{-t/\tau_r} \Theta(t)$ (Eq. 4), plus a constant initial threshold $c$, and $\tau_r$ indicates neuron time constant for the refractory kernel.

This operational principle of the neuron is illustrated in Fig. 2a. In contrast to LIF neurons, which reset their membrane potential to a resting state upon reaching the threshold (Fig. 2a, upper), thereby losing historical information accumulated on the membrane potential, SRM neurons maintain a continuous membrane potential trace. The SRM neurons employ two kernels, $\varepsilon$ and $\upsilon$, each with distinct time constants ($\tau_s, \tau_r$) for response and refractoriness respectively (Fig. 2a, lower). The interaction between these two processes operating on different timescales engenders a rich repertoire of spiking patterns, including bursting and phasic spiking, which are absent in the simpler LIF neurons. The exponential decaying refractory kernel introduces a second forgetting mechanism by adaptively elevating and then decaying the firing threshold, like the forget gate in the long short-term memory (LSTM) models. This adaptive threshold lets each neuron selectively retain the memory of its most recent spike while attenuating the influence of earlier firings. This mechanism is missing in the LIF neurons and potentially translates into enhanced accuracy over the LIF neurons in tasks that require the processing of complex temporal patterns.

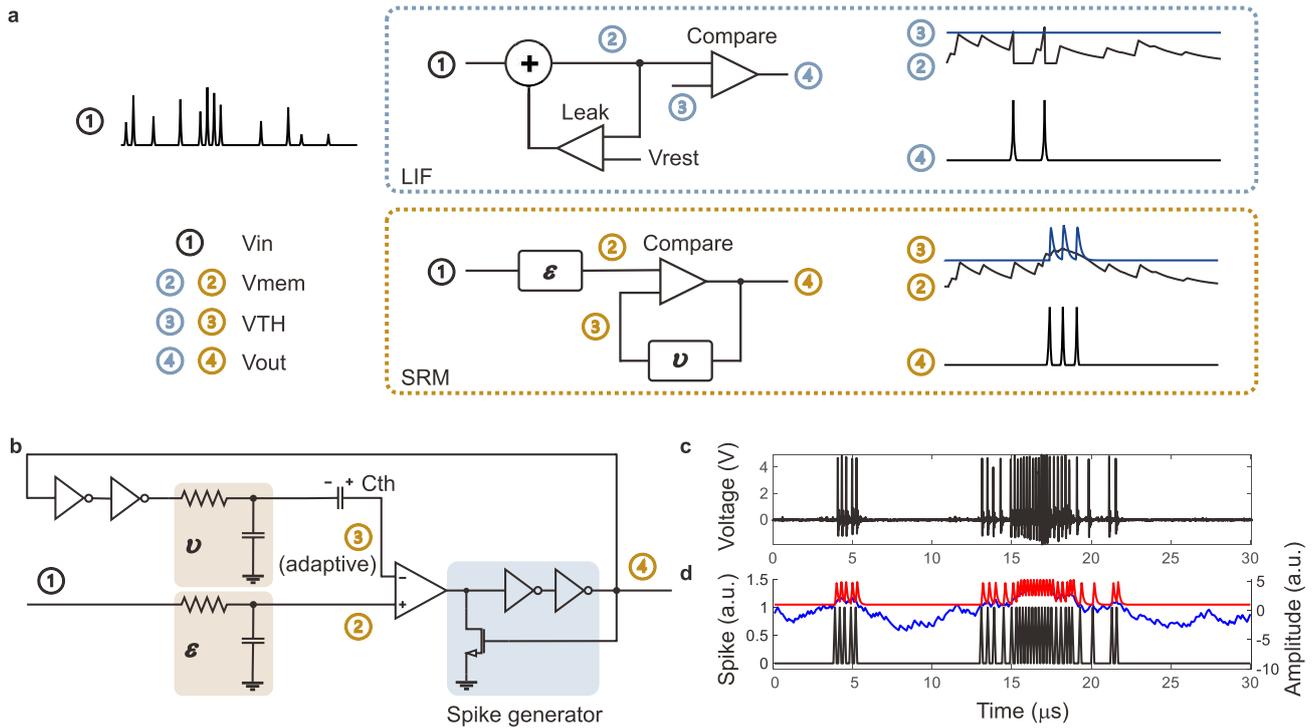

**Fig. 2 | Operation of SRM neurons, contrasted with LIF neurons.** (a) SRM neurons (lower) integrate input spikes ("Vin") using an $\varepsilon$ kernel to generate membrane potential ("Vmem") and dynamically adjust the firing threshold ("VTH") via a $\upsilon$ kernel to modulate firing activity ("Vout"). These two kernels enable the selective retention of recent information and the attenuation of older influences within the neuron. Unlike LIF neurons (upper), SRM neurons do not reset the membrane potential to a resting state ("Vrest"), enabling more diverse spiking patterns and extended temporal dependence, which potentially enhances inference accuracy. This also reduces the latency associated with the resetting process in LIF neurons. (b) Schematics of the core soma circuit of the SRM neuron. Panels (c, d) show representative SRM responses to a 30-μs post-synaptic spike train. Panel c displays the measured response while panel d presents the corresponding calculated response (output spikes in black, left axis; membrane potential in blue and dynamic threshold in red, right axis). Pads and external probes were not attached to the membrane potential and the dynamic threshold nodes to avoid parasitic capacitance and/or resistance, ensuring the integrity of neuron dynamics.

The CMOS realization of the custom-designed SRM neuron function is shown in Fig. 2b, which primarily compromises two passive RC filters (implementing $\varepsilon$ and $\upsilon$ kernels) and a comparator. Let's take the $\varepsilon$-kernel that generates the membrane potential as an example. When an input post-synaptic voltage spike ("Vin", representing $\boldsymbol{o}^{(l)}(t)$ in Eq. 1) arrives, it injects charge onto the capacitor (C) through the resistor (R). Consider the input spike as an impulse, the capacitor voltage rises rapidly. Subsequently, in the absence of further input (i.e., when the input returns to baseline), the capacitor discharges through the resistor. The discharge process is exponential with the voltage across the capacitor following $e^{-t/\tau_s}$, where the time constant $\tau_s = R_s C_s$. This natural exponential response of the RC circuit to an impulsive input forms the basis of the kernel $\varepsilon(t) = \frac{1}{\tau_s} e^{-t/\tau_s} \Theta(t)$ (More details in Methods). When multiple input spikes arrive, the individual exponential post-synaptic potentials generated by each spike simply sum up (superimpose) on the capacitor because of the linearity of the RC filter. This linear superposition is precisely what the convolution operation in Eq. 2 describes, making the RC circuit an elegant analog means of performing this computation. The continuous voltage across the capacitor ("Vmem") therefore accurately represents the convolution of the entire train of input spikes with this exponential kernel. Similarly, the $\upsilon$-filter, with its own time constant, implements the exponential refractory kernel through the same physical principles. It is noteworthy that, although it is feasible to design more complex temporal kernels using other configurations of passive circuit components, the simple first-order RC filter, yielding an exponential kernel, already provides an effective and resource-efficient solution for the SRM dynamics we aim to capture. Its performance, as demonstrated by our results described in the following sections, indicates its sufficiency for complex temporal processing tasks.

A pair of TIAs converts incoming differential current spikes into input voltage spikes ("$V_{in}$"), which represent $\boldsymbol{s}^{(l)}(t) = \sum \delta(t - t_i)$, as defined in Eq. 1. Next, an RC filter integrates these input spikes through the natural convolution operation described above to produce the membrane potential "$V_{mem}$". When $V_{mem}$ exceeds the adaptive threshold "$V_{th}$", an output spike "$V_{out}$" is generated by the comparator. This output spike also feeds back to the $\upsilon$ filter, adaptively increasing the threshold "$V_{th}$" which was initially set to "$c$", implementing refractoriness. The inverters placed before and after $V_{out}$ shape the output spikes and isolate the feedback $\upsilon$ filter, respectively. Experimental validation of neuronal responses was conducted by feeding a fabricated on-chip neuron with post-synaptic spike trains of 30 µs duration at a temporal resolution of 100 ns. Fig. 2c presents a representative measured response. Between 15 and 19 µs, the neuron exhibited rapid, clustered spikes followed by a clearer period of quiescence, characteristic of a bursting pattern because of the adaptive threshold that is absent in LIF neurons. This measured response aligns closely with the neuronal behavior predicted by Eqs. 2-4 (Fig. 2d).

A notable feature of our hardware SRM neuron design, leveraging exponential kernels implemented with analog RC filters, is its proportional time-scaling behavior. If the input spike trains are time-scaled by a factor $\alpha$, and if the time constants of the kernels ($\tau_s$, $\tau_r$) are correspondingly scaled by the same factor $\alpha$, (achieved by selecting appropriate R and C values for the filters), the output spike train will also be a time-scaled version of the original without changing the amplitude (see detailed mathematical proofs in Methods). If we were to implement a system to process spike signals at their original, often slower, biological timescales, the RC values required to realize the necessary time constants would be substantial, leading to a large chip area for the capacitors and hindering large-scale integration. However, given the proportional time-scaling property described, we can implement compact, area-efficient RC filters with much smaller time constants. For example, using the mentioned 2 pF and 700 fF capacitors enables microsecond-scale operations, and even smaller components can be made with more advanced technology nodes. This design allows the system to process input data that has been significantly time-compressed (accelerated), thereby drastically reducing processing latency for event-based data or other spike-formatted signals, while maintaining the essential computational dynamics of the SRM neurons.

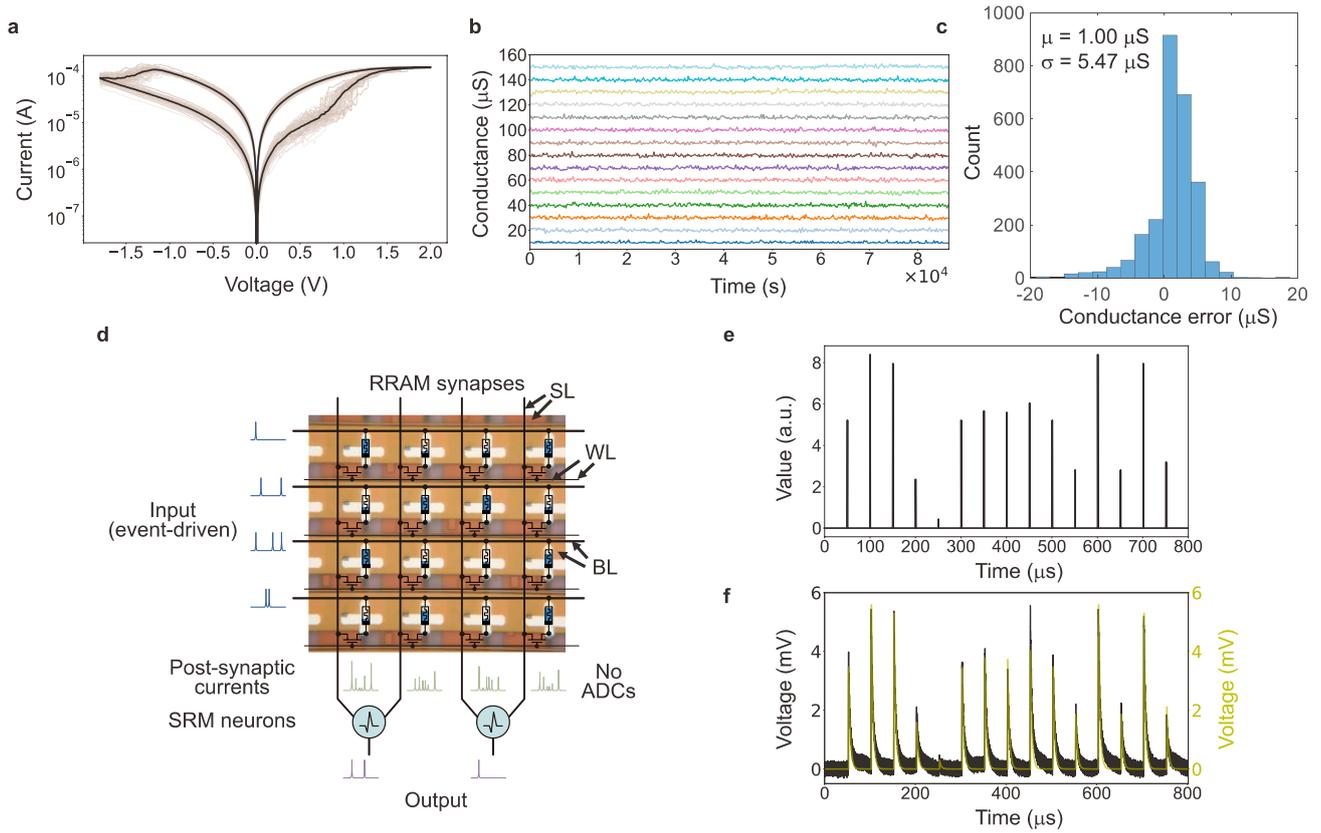

**Fig. 3 | Experimental characterization of the RRAM synapse network.** (a) Switching characteristics of a single RRAM device over 50 cycles of DC sweeps. The black line represents the average switching characteristics. (b) Retention and read-disturb characteristics of the device states with 0.2-V read pulses over a 24-hour period, using 15 RRAM devices, each programmed to a distinct conductance state. The 15 states are evenly separated by 10 µS within the range 10~150 µS. (c) Distribution of RRAM conductance programming errors after deploying an offline-trained model on a PE. (d) Hybrid RRAM-CMOS structure implementing a network layer, where an RRAM crossbar array serves as a synapse network, with every two columns connected to an SRM neuron for processing post-synaptic currents into output voltage spike trains. (e, f) Experimental validation of synaptic computation showing that measured post-synaptic current, converted to voltage (f, black), corresponded with the calculated result of matrix-vector multiplication of input spike sequences and synaptic weights (e). The experiment applied 0.2-V input spikes to the RRAM crossbar array and converted the post-synaptic current to voltage via an off-chip grounded resistor.

## Integration of memristor synapse network with custom-designed SRM neurons

Our custom SRM neurons are directly integrated with the 1T1M memristor crossbar array as the synapse network. The resistive switching behavior of our on-chip memristor devices is shown in Fig. 3a, with a set voltage of around 1 V and reset voltage of around -1.5 V. The memristor can be reliably programmed into multiple states, with Fig. 3b shows 15 evenly distributed conductance states within 150 µS that retained for 24-hour period with no noticeable drift. Subsequently, we deployed the network weights of an offline-trained SNN model to the memristor crossbar using an iterative write-verify method with a writing tolerance of ±5 µS. The resultant readout conductance errors (Fig. 3c) show a standard deviation of 5.47 µS.

Functionally, the memristor array acts as a synaptic matrix within our SNN, where each SRM neuron processes the weighted post-synaptic currents from two differential columns (Fig. 3d). Different from the crossbar arrays for general matrix multiplication, both the input and output signals are in the form of voltage spikes instead of steady voltage signals. To experimentally demonstrate this synaptic computation, input spike sequences (Fig. S3a), generated by an AWG, were applied to the rows of memristor array that have been programmed to represent the SNN weights (experimentally readout conductance map after programming and target conductance map are shown in Fig. S3b and S3c, respectively). To measure the post-synaptic current, we connected a column output to the ground through an external sense resistor. The voltage drop across this resistor was then measured as a proxy for the current (Fig. 3f, black trace). This specific configuration is used only when performing this measurement. The measured output result closely matches the calculated result (Fig. 3e) from the matrix-vector multiplication of input spikes and synaptic weights. The observed slow decay process in the measured result is due to the parasitic capacitance within the off-chip circuits and the connected resistor used. To verify this, a simulation incorporating this parasitic capacitance (Fig. 3f, yellow) demonstrates consistency with the measured result.

Our memristor synapses and custom SRM neurons are integrated directly and signals are transmitted in the analog domain, without time- and energy-intensive analog-digital data conversion steps. Also, once synaptic weights and neuronal thresholds are initialized, the system operates asynchronously, driven solely by input spike events that are processed individually upon arrival, without any system-level clocking, buffering, or batching into frames. By avoiding such synchronization, the system directly processes the original temporal stream of data, leveraging its native temporal information, and eliminates the power and latency overhead associated with clock management and frame generation.

Unlike previous SRM-based SNNs [36, 38], which model each synapse response with temporal kernels, our approach delegates complex temporal dynamics exclusively to the neurons. This design choice is crucial for energy efficiency inherent to memristor crossbar arrays and SNN systems: implementing a kernel for each synapse response would otherwise generate a continuous voltage waveform input to the memristor synapse network, negating the zero static energy consumption benefit of memristors in the absence of input spikes and undermining the event-driven nature of SNNs.

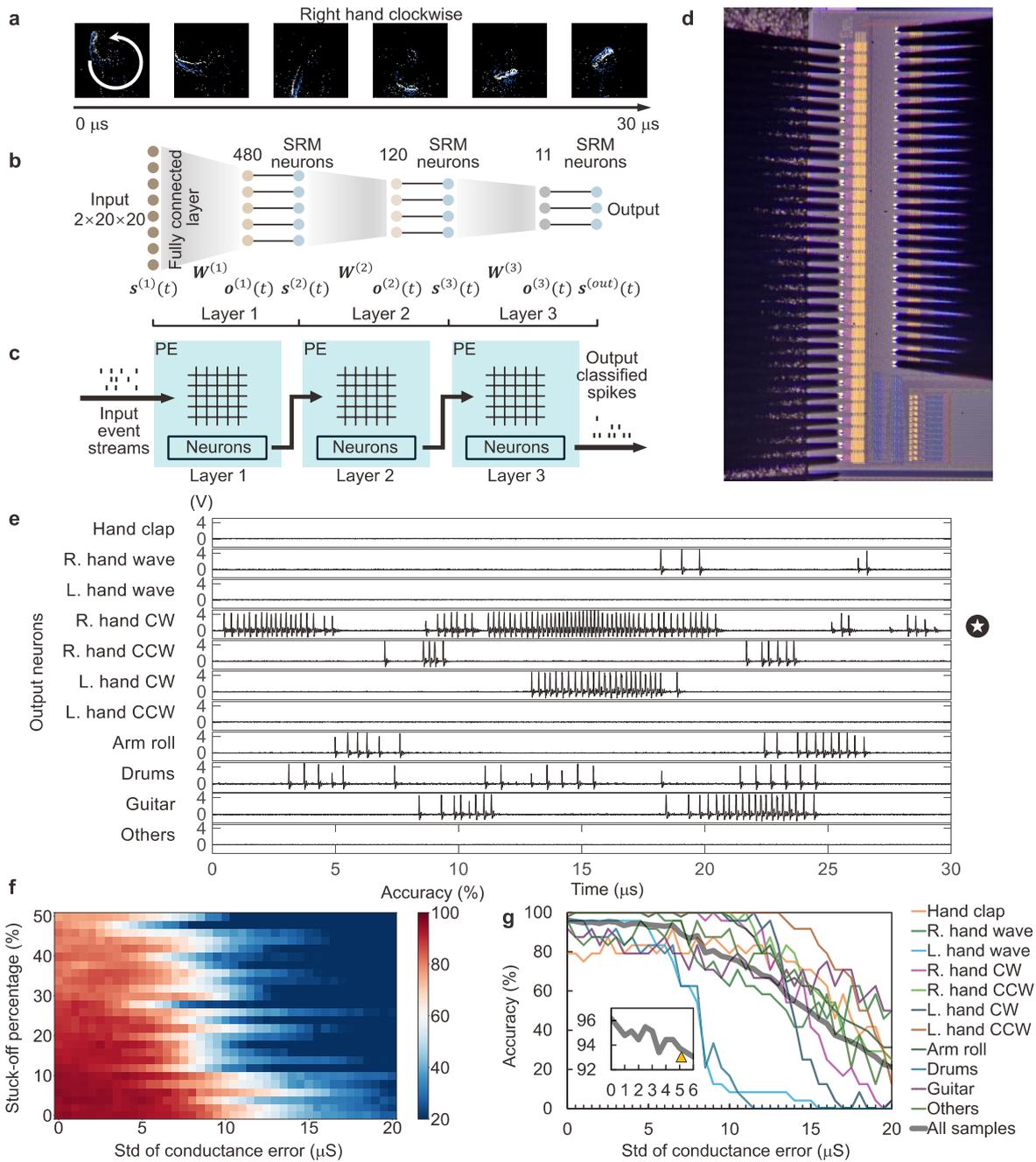

**Fig. 4 | High-speed gesture recognition experiments.** (a) A sample from the DVS128 Gesture dataset, accelerated to a duration of 30 μs, showing a right-hand clockwise waving gesture. (b) A three-layer fully connected SNN architecture, trained offline to classify gestures from the DVS128 Gesture dataset. (c) System diagram illustrating the mapping of each network layer onto a PE macro. (d) Micrograph of a PE macro, interfaced with external measurement circuits via probes. (e) Measured output-layer spike trains during classification of the gesture in (a); the neuron corresponding to the target class exhibits the highest spike count. (f) Simulated inference accuracy degrades with the increase of stuck-off devices in the crossbar array and conductance-error standard deviation (Std), while (g) shows detailed accuracy degradation at a fixed 6% stuck-off percentage: accuracy remained above 93% as the conductance-error Std increases to 6 μS (inset), with the yellow marker indicating measured accuracy at the readout conductance-error Std.

# High-speed spatiotemporal pattern recognition

To demonstrate the high-speed inference capabilities of our SNN system on complex spatiotemporal data, we performed gesture classification experiments using the DVS128 Gesture dataset (Fig. 4a). This dataset comprises approximately 6-second neuromorphic event recordings of 11 distinct hand and arm gestures [39], which we split into 1,176 training samples and 288 test samples. A three-layer fully connected SNN (480-120-11 neurons) was implemented (Fig. 4b). It processes a 128×128×2 input that undergoes pre-processing via 6-pixel padding and 7×7 pooling. Our architecture maps each network layer to a memristor-synapse-CMOS-neuron SNN macro (Fig. 4c), with the last layer experimentally implemented in our physical chip, while other layers are simulated using a hardware-calibrated model. Fig. 4d illustrates our SNN macro interfacing with external measurement circuits via probes touching the chip pads.

The SNN was trained offline using a surrogate-gradient backpropagation algorithm [40, 41], enabling direct training with temporal spike dynamics and the custom neuron response. A discussion of training SNNs with backpropagation through surrogate gradients can be found in the Supplementary Note S1. This direct training can achieve high accuracy without the prolonged inference time associated with ANN-to-SNN conversion [42]. The neuron time constants ($\tau_s$, $\tau_r$) are critical hyperparameters for a better performance, as they need to align with the temporal scale of a given task. In addition, the choice of the refractory kernel time constant $\tau_r$ with respect to that of response kernel $\tau_s$ regulates the "forgetting" mechanism and also influences the accuracy (see detailed discussion in Fig. S2). By carefully selecting these time constants ($\tau_s$ = 50 ms, $\tau_r$ = 5 ms) based on the time scale of the spikes in the training dataset (5-ms resolution), our network model achieved a SOTA testing accuracy of 97.71±0.47% with 443 K parameters as compared to the best reported accuracy of 98.0% with 1,100 K parameters [43]. More training details can be found in the Methods section, and a benchmark comparison of our algorithm's accuracy with recent reports on the DVS128 Gesture dataset is presented in Supplementary Table S1.

Because of the adaptive threshold and adjustable refractory mechanism, the SNNs with SRM neurons show a clear benefit in model performance compared to those with LIF neurons. To evaluate this, we conducted a comparative algorithmic experiment. By substituting SRM neurons with LIF neurons within an otherwise identical network structure (Fig. 4b), we retrained the modified SNN model with the same loss function and training procedure on the DVS128 Gesture dataset using snnTorch [4], incorporating learnable decay rates for each LIF neuron. The LIF-based model attained a maximum accuracy of 94.44% (Fig. S12) and an average accuracy of 92.71±1.18% across five independent trials. In contrast, the SRM-based model achieved a peak accuracy of 98.26% and an average accuracy of 97.71±0.47%, suggesting the potential performance advantages of SRM neurons, a benefit that could be enlarged for tasks requiring the processing of more complex temporal patterns.

For spatiotemporal pattern recognition inference in hardware, learned weights were mapped and programmed onto memristors. Given our limited number of devices in our proof-of-concept demo chip, the first two layers of the classification were emulated based on measurement, including 5.53% stuck-off devices, while the weight values of the last layer were experimentally readout from our chip (Fig. S4 experimental conductance map after programming, see *Methods* for measurement system setup details). During the inference phase, all the on-chip controllers were disabled via clock gating, making the memristor chip operate in an event-driven mode that was solely triggered by input spikes. The resulting output spike trains produced by the neurons were captured by an FPGA and classification is determined by the output neuron that exhibits the highest spike count within the final layer. This winner-take-all approach is inherently resilient to variations in spike timing due to fabrication mismatches and circuit noises. The classification outcome depends on which neuron is most active over the entire inference process, rather than the precise timing of individual spikes, making the system robust against timing variations. Figs. S5 and S6 provide a detailed illustration of this workflow, and the experimental setup used to measure inference accuracy, respectively.

Implementing the millisecond-scale time constants used during training would require large, area demanding, resistors and capacitors (e.g., ~100 kΩ resistor and ~10 nF capacitor for 1 ms time constant), making the solution unscalable. However, due to the proportional time-scaling behavior of our neuron, our design allows us to scale down the time constants to achieve the exact same behavior, enabling the use of smaller on-chip components and faster processing of spiking signals. Given the RC for implementing the kernel need to be significantly larger than the parasitic for accurate computing, we chose conservative values of $\tau_s$=1 μs and $\tau_r$=100 ns (using R=500 kΩ/142.857 kΩ, C=2 pF/700 fF) for our demonstration, but they can be further scaled with a more aggressive design or using more advanced technology node in future studies. Since these time constants are 50,000 smaller than those used during training, we accelerated the input signals by the same factor. Consequently, each gesture sample, originally 1.5 seconds long, was processed in just 30 μs (a 50,000× speed up). As shown in Fig. 4e, for a right-hand clockwise waving gesture (Fig. 4a), the target neuron corresponding to that gesture exhibited the highest spike count, indicating a correct classification.

Experimentally, our hardware SNN system achieved a classification accuracy of 93.06% on the DVS128 Gesture dataset (classification confusion matrix shown in Fig. S7b). The 5.20% accuracy reduction compared to the software baseline is attributed to non-idealities in the lab-fabricated memristor synapses and neuron circuits. To better understand this degradation, we replaced the memristor synapses and the neuron circuits individually with their ideal software counterparts to isolate the effects of each hardware imperfection. Fig. S8 indicates that both the memristor synapse variation and the neuron circuit non-idealities contributed to inference errors: non-ideal neuron circuits lower accuracy by an average of 3.19%, while RRAM-synapse variability causes an average accuracy drop of 3.43%, when compared to the software baseline. We further modeled memristor conductance variation using Gaussian normal distributions and introduced a controlled percentage of "stuck-off" devices, defined as devices with random conductance values below empirically derived 4 µS. Using the model, it is shown that inference accuracy declines with increasing conductance error and stuck-off percentage, but the system shows resilience to these errors. The system maintains an accuracy of over 85% with a stuck-off rate within 10% and a conductance error standard deviation up to 6 µS (Fig. 4f). At an experimentally observed stuck-off rate of 6%, the accuracy climbs above 93% with the conductance error standard deviation capped at 6 µS (Fig. 4g). While the inherent conductance variation of memristors is unlikely to decrease due to physical limitations, the stuck-off ratio is expected to be significantly lower in devices from an industrial foundry. Our own best lab-made chip already demonstrated a 99.8% yield [9]; at this high level, a conductance variation as large as 6 µS trims the accuracy by only 2.78%.

To demonstrate versatility, the system was also tested on Neuromorphic MNIST (N-MNIST) [44] dataset. Because the N-MNIST's 34×34×2 input size is different from that of DVS128 Gesture dataset, we adapted the input pre-processing with a 3-pixel padding and 2 × 2 pooling. This allows us to maintain the network architecture (with the output layer adjusted to 10). Retrained with time constants of $\tau_s$ =10 ms and $\tau_r$ =1 ms, the model achieved a software accuracy of 98.29%. Subsequently, the learned parameters of the last layer were programmed onto the devices (Fig. S9 depicts the distribution of readout conductance error in the final layer), and the input spikes were accelerated by a factor of $10^4$ relative to the original dataset and training. The experimentally measured inference accuracy was 94.73% against the software baseline of 98.29% (Fig. S10). This demonstrates the adaptability to different tasks without hardware modification.

# Table 1 | Benchmark comparison of the system with recent SNN systems

| | ISSCC'23 [45] | ISSCC'22 [51] | ISSCC'21 [46] | ISSCC'21 [47] | JSSCC'23 [48] | ISSCC'23 [49] | ISSCC'23 [50] | IEDM'19 [14] | This work |
|---|---|---|---|---|---|---|---|---|---|
| Technology | 40 nm | 180 nm | 65 nm | 180 nm | 28 nm | 28 nm | 28 nm | 130 nm | 180 nm |
| Network | SNN + CNN | SNN | SNN | SNN + CNN | SNN | SNN | Spiking RNN | SNN | SNN |
| Synapse Operation and Memory Type | CNN: Analog, RRAM SNN: Digital, SRAM | Digital SRAM | Digital - | Digital - | Digital - | Digital SRAM | Digital SRAM | Analog RRAM | Analog RRAM |
| Synapse Weight Precision (bit) | CNN: Analog SNN: 1 | 4 | 1 | 8 | 8 | 8 | 8 | 3 | Analog |
| Neuron Operation | Digital | Analog | Digital | Digital | Digital | Digital | Digital | Analog | Analog |
| Dataset | Public UAV dataset | MIT-BIH Arrhythmia | HeySnips & GSCD | MIT-BIH Arrhythmia & GSCD | LFW | NMNIST, DVS Gesture, N-TIDIGIT & SeNic | DVS Gesture, Spiking Heidelberg Digits & Delayed cue integration | MNIST | DVS Gesture & NMNIST |
| Accuracy | - | 90.5% | 96.5%@20dB SNR | 99%@ECG 94%@KWS 1word | 96% | 96.0%@NMNIST, 92.0%@DVS Gest.(10classes, the entire sample), 92.6%@KWS 1word, 95.7%@SeNic | 87.3%@DVS Gest.(10classes, max-8s sample), 90.7%@KWS 1word, 96.4%@Nav.(2decision) | 84% | **93.06%@DVS Gest.(11classes,1.5s sample), 94.73%@NMNIST** |
| Latency | 20 ms | 40 μs | 100 ms | 348 ms | 681 μs | - | - | - | < 14.31 μs |
| Power (infer.) | 8.71 mW | <350 nW | <570 nW | 378 nW@KWS 1.68 μW@ECG | 63.2 μW [a] | 94.1 μW [a c] | 80 μW@13MHz | 1.5 mW | 43.83 mW [a b] |
| Energy per Sample (infer.) | - | - | - | - | - | 343nJ@NMNIST, 3.9μJ@DVS Gest., 6.1μJ@KWS, 582nJ@SeNic | 46.13μJ@DVS Gest., 4.37μJ@KWS, 1.35μJ@Nav. | 24.48 nJ | 1.94μJ@DVS Gest. 1.78μJ@NMNIST |
| Energy Efficiency [d] | 36.77 TSOPS/W | 1.89 TSOPS/W | 0.67 TSOPS/W | - | 75.05 TSOPS/W | 0.67 TSOPS/W | 0.19 TSOPS/W | 0.28 TSOPS/W | **101.05 [a b] TSOPS/W** |
| CIM | ✓ | ✓ | ✗ | ✗ | ✗ | ✗ | ✗ | ✓ | ✓ |
| Direct Temporal Stream Processing | ✗ | ✗ | ✗ | ✗ | ✗ | ✓ | ✓ | ✗ | ✓ |
| Full Analog | ✗ | ✗ | ✗ | ✗ | ✗ | ✗ | ✗ | ✗ | ✓ |
| w/o Frame | ✗ | ✓ | ✗ | ✗ | ✗ | ✓ | ✓ | ✓ | ✓ |
| Event Driven at Circuit Level | ✗ | ✓ | ✗ | ✗ | ✗ | ✗ | ✗ | ✗ | ✓ |

[a] Static power consumption is included. [b] Power of chip I/Os is included. [c] Calculated based on energy per sample and sample duration. [d] For consistency, each synaptic operation counted as 2 operations.

## Performance benchmark and scalability discussion

In our asynchronous, event-driven system, the memristor crossbars and analog neuron circuits process the spike signals in continuous real-time with minimum latency. The total processing latency therefore is only attributed to the signal propagation delay (in the connection wire and memristor crossbars) and the neuron circuit latency. Post-layout simulation using Cadence Virtuoso indicates a total signal propagation latency of approximate 150 ps within our crossbar. As this is five orders of magnitude less than the duration of input spike streams, its impact is negligible. Our neuron circuit, however, causes non-negligible latency, which is measured at 14.31 μs for the combined three-layer circuit for each 30 μs input sample (see Methods for measurement details). Therefore, the total inference time is the sum of the input duration and this circuit latency, which leads to 44.31 μs to classify a gesture in the DVS dataset. Notably, there is significant potential for improvement. The current implementation, which uses a 180 nm technology node, did not optimize for performance. More advanced nodes and optimized designs would considerably reduce this latency.

The ability to process accelerated spike signals also promises significant benefit in energy efficiency. The asynchronous, event-driven system paired with the non-volatile memristor devices consumes energy only when there is a spike. We evaluated the consumed power and energy of our system for DVS128 Gesture dataset inference by combining experimental data and simulation results, and obtained a total average power of 43.83 mW during inference, resulting in an energy expenditure of 1.94 μJ per sample. This translates to a performance of 4.43 TSOPS (tera-synaptic operations per second) and an energy efficiency of 101.05 TSOPS/W. It is noteworthy that there is significant potential for further improvement, as the current number is based on our experiments that take I/O and off-chip connection into consideration. A breakdown analysis in Fig. S13 shows that I/O is the primary consumer of chip area and a major contributor to energy usage. The current design measures spike outputs off-chip, which limits performance. A fully on-chip integrated version would bypass this bottleneck, suggesting that a much higher level of performance and efficiency could be realized (see Supplementary Note S2). Table 1 provides a comparative analysis of our system with recent SOTA SNN systems [14, 45-51], which shows that our system is highly competitive, even with the overhead of off-chip I/O considered.

The ultimate energy efficiency is determined by the bandwidth of the passive memristor crossbar— that is, the shortest spike it can support. If input spikes cannot be simply accelerated further, efficiency can still be improved by shortening the duration of each spike. To explore this limit, we analyzed the minimum spike width that our crossbar array can support, which is defined as the full width of an input spike resulting in a 70% attenuation of the output voltage from an ideal differential TIA pair. Post-layout simulations indicate that our crossbar array can accommodate spikes as narrow as 267 ps (Fig. S11b), considering all parasitic and oversized selector transistors in 180 nm technology. The value can still be significantly improved by better design and more advanced technology node. Reducing the input spike width to this minimum only marginally impacts the inference accuracy of our SNN model on the DVS128 Gesture dataset, causing only a minor decrease from 98.26% to 97.92% (Fig. S11c). This demonstrates a possible pathway to substantial improvements in latency and energy efficiency to 38.21 ns and 18.03 PSOPS/W, which could be unlocked in the future with the development of optimized, high-speed neuron circuits in advanced technology nodes.

## Conclusions

In summary, this study presents and experimentally demonstrates a fully integrated memristive SNN, designed for low-latency, energy-efficient inference of spatiotemporal patterns in microsecond-scale data streams at the edge. The system, implemented with a chip of 180-nm technology node, performs direct stream processing on each accelerated 30-μs DVS128 Gesture dataset sample within 44.31 μs, achieving an accuracy of 93.06% compared to the 98.26% software baseline. Despite the modest accuracy reduction, the system's full-analog physical computation and direct temporal stream inference paradigm results in a measured 2.80× reduction in latency and a 1.35× enhancement in energy efficiency compared to the prior SOTA implementations. These characteristics establish the system as a prime candidate for integration with high-speed sensing devices, such as dynamic vision sensors operating at microsecond-level temporal resolutions, thereby advancing ultra-fast, in-sensor machine vision and other rapid-response edge computing applications.

## Methods

**Overview of the PE macro and its measurement board**

The PE macro implements individual SNN layers. As illustrated in Fig. 1b, the PE integrates a 128×24 RRAM crossbar array as synaptic network together with 12 CMOS neurons to process the synaptic outputs. Each synaptic connection is realized by a differential pair of Ta/TaOx/Pt RRAM devices configured in a one-transistor-one-memristor (1T1M) structure (dotted outline in Fig. 1d). Signed synaptic weights are encoded in the conductance difference between the "positive" and "negative" devices of the pair.

During inference, incoming spike events are applied to the horizontal bit-lines of the crossbar array, while the horizontal word-lines are set at the VDD voltage to turn on all access transistors. The resulting weighted currents flow vertically along the source-lines, where each neuron employs a differential trans-impedance amplifier pair to convert the current into a post-synaptic voltage ("Vin"). However, in our current proof-of-concept chip, the memristor array's inputs share I/O resources that are managed by the controllers for routing selection. The operational speed of the controllers creates a bottleneck, precluding the high-speed signal input. To circumvent this hardware limitation and validate the neuron's high-speed processing capabilities, we employed a two-step experimental methodology. First, we experimentally read out the synaptic conductance values directly from the programmed physical memristor crossbar. These measured values capture the experimental non-idealities (device-to-device variations and stuck-off devices) of the synaptic array. Second, these conductance values were used to generate the corresponding post-synaptic voltage signals, which were fed to the on-chip integrated neurons via the high-speed AWG. This methodology provides a focused, proof-of-concept demonstration of the neuron's high-speed spike generation capabilities, which are the primary subject of this investigation and would otherwise be obscured by the speed constraints of the current array controllers. Future iterations with on-chip integrated PEs will eliminate the need for shared I/O and routing controllers, thereby removing this bottleneck and enabling direct, high-speed signal input to the synaptic array.

The physical layout is optimized for low wire resistance: each 1T1M unit cell measures 30 μm × 30 μm, and both bit-lines and source-lines are widened to 20 μm. Each neuron occupies 49 μm × 361 μm (17,689 μm²), of which the TIA pair accounts for 13,671 μm² and the soma circuit occupies the remaining 4,018 μm².

The die is interfaced with a custom probe card (Fig. 1j) that plugs into the measurement board (Fig. 1i) via coax ribbon cables. The board houses an FPGA that controls the chip and captures neuron outputs to the host computer.

**Time-scaling effect of the system**

The temporal invariance of the synaptic network allows us to focus on the neuron's behavior to elucidate the system's time-scaling properties. The convolution operation of the response kernel $\varepsilon$ (Eq. 2), which yields the membrane potential $\boldsymbol{u}^{(l)}(t)$, can be expressed as

$$\boldsymbol{u}^{(l)}(t) = \boldsymbol{o}^{(l)}(t) * \varepsilon. \tag{7}$$

Applying the Laplace transform to Eq. 7 yields:

$$\boldsymbol{U}(s) = \boldsymbol{O}(s)H(s), \tag{8}$$

where $\boldsymbol{O}(s)$ and $\boldsymbol{U}(s)$ represent the Laplace transforms of post-synaptic voltage spike trains $\boldsymbol{o}^{(l)}(t)$ and membrane potential $\boldsymbol{u}^{(l)}(t)$, respectively. The transfer function $H(s)$, corresponding to the response kernel with time constant $\tau_s$, is given by

$$H(s) = \frac{1}{\tau_s s + 1}. \tag{9}$$

Scaling temporally the input spike trains $\boldsymbol{s}^{(l)}(t)$ in Eq. 1 by a factor $a$, denoted as $\boldsymbol{s}^{(l)}(at)$, results in corresponding time-scaled post-synaptic voltage spike trains $\boldsymbol{o}^{(l)}(at)$, with its Laplace transform changed to $\boldsymbol{O}'(s) = \frac{1}{a}\boldsymbol{O}(\frac{s}{a})$. By scaling the response kernel's time constant by $1/a$, the transfer function changes to

$$H'(s) = \frac{1}{\frac{\tau_s}{a}s+1} = H\left(\frac{s}{a}\right), \tag{10}$$

without altering the amplitude response. Consequently, the membrane potential in the Laplace domain becomes:

$$\boldsymbol{U}'(s) = \boldsymbol{O}'(s)H'(s) = \frac{1}{a}\boldsymbol{O}\left(\frac{s}{a}\right)H\left(\frac{s}{a}\right) = \frac{1}{a}\boldsymbol{U}\left(\frac{s}{a}\right), \tag{11}$$

yielding a membrane potential $\boldsymbol{u}^{(l)}(at)$, which is temporally scaled by the same factor $a$ while preserving the response characteristics. Similarly, scaling the refractory kernel $\upsilon$'s time constant by $1/a$ results in the dynamic threshold temporally scaled by the same factor $a$. Therefore, scaling both kernel time constants in proportion to the input spike trains' temporal scaling factor maintains consistent neuronal and systemic response characteristics without altering the amplitude response. This property allows our system to achieve inference with reduced latency.

**Time-constant adjustment mechanism of the system**

The transfer function associated with the response kernel $\varepsilon$, characterized by time constant $\tau_s$, (Eq. 9), undergoes a transformation when subjected to an input signal time-scaled by a factor $a$, denoted as $\boldsymbol{o}^{(l)}(at)$. This transformation is represented by:

$$H(s)|_{\tau=\tau_s} = \frac{1}{\tau_s s+1} = \frac{1}{a\tau_s \frac{s}{a}+1} = H\left(\frac{s}{a}\right)\Big|_{\tau=a\tau_s}, \tag{12}$$

which demonstrates an effective scaling of the time constant by a factor of $a$ without any amplitude scaling. This effective scaling of the time constant also applies to the refractory kernel $\upsilon$.

Therefore, temporal scaling of input spike trains effectively scales the time constants of our system's transfer function, while having no amplitude scaling. This allows our system to adaptively align neuronal time constants to the specific task, resulting in competitive inference accuracy and the flexibility to handle different tasks.

**Constrained SNN training for optimized hardware implementation**

To reduce the discrepancy in inference accuracy between the network model and its hardware implementation, we employed weight clipping (restricted to [-5, 5] for the initial two layers and [-3, 3] for the final layer) and activation clipping (confined to [0, 5]) during training. These measures enforce the constraints representative of hardware systems and improve algorithm performance by introducing non-linearity. The initial neuron threshold was fixed to be 1. Additionally, we applied data augmentation techniques (including random cropping, perspective transform and rotation on the input samples) and L2 regularization (parameterized by $\lambda = 5/(2*N)$, where $N$ represents the number of training samples) to enhance training effectiveness, resulting in a SOTA algorithm accuracy. L2 regularization also helps strengthen the model's resilience to weight noise by promoting smaller weights. This makes the variations in RRAM device conductance, which represent synaptic weights, have a relatively smaller effect on inference accuracy.

**Latency measurement**

Figure S15 illustrates the latency incurred by a single layer in the three-layer SNN system. To benchmark the inference latency, we measured the first-spike latency and the latency of a full spike train under a worst-case scenario. The first-spike latency is defined as the interval between the onset of the input stimulus and the occurrence of the first spike in response to that stimulus. By injecting a saturating step voltage to a representative on-chip neuron to induce immediate firing, we measured the first-spike latency to be 0.15 μs.

During the inference for the DVS128 Gesture and NMNIST dataset samples, a continuously firing neuron can produce up to 300 spikes. By maintaining the injected voltage at the saturating level to force the neuron to fire continuously, the latency to emit these 300 spikes was measured as 4.77 μs. The actual number of spikes a neuron generates during inference is significantly lower. During the inference of DVS128 Gesture and NMNIST datasets, the maximum number of output spikes by a neuron was 149 and 134, respectively.

The total latency for a single layer ("layer latency") is determined by the entire spike train. The first-spike latency of 0.15 μs is the initiation of this process. Therefore, the worst-case per-layer latency is conservatively bounded by the latency to generate the maximum possible spike train, which we measured as 4.77 μs. Propagating this latency across the three-layer network yields a total worst-case inference latency of 14.31 μs (4.77 μs × 3) per dynamic sample. As real inference produces far fewer spikes than the worst case (⩽149 for DVS128 Gesture dataset and ⩽134 for NMNIST dataset), practical latencies are appreciably lower than this conservative ceiling.

# Data availability

The data that support the findings of this study are included within the main text and Supplementary Information and are also available from the corresponding author upon reasonable request.

# Code availability

A demonstration code of the SNN model for gesture classification on the DVS128 Gesture dataset is available in the public GitHub repository:

https://github.com/wzhku/SNN_for_High_Speed_Processing

# Acknowledgements

The work described in this paper was partially supported by a grant from the Research Grants Council of the Hong Kong Special Administrative Region, China (Project No. HKU 27210321, HKU 17207925,

HKU C7003-24Y, HKU C1009-22GF, HKU T45-701/22-R), National Natural Science Foundation of China (62122005), ACCESS—AI Chip Center for Emerging Smart Systems — sponsored by InnoHK funding, Hong Kong SAR, and Croucher Foundation.

## Author contributions

Z.W., S.W., and C.L. conceived the idea. Z.W. developed the SNN models. Z.W., S.W., H. K.-H. S., and C.L. designed the neuron circuit and chip architecture. Z.W., S.W., R.M., and Y.X. designed the 1T1M synapse array and implemented physical layout of the chip. Z.D. developed and optimized the memristor devices and integrated them with the CMOS. Z.W. developed the test system. Z.W. and S.W. developed the software toolchain and conducted all chip measurements. C.L. and P.L. supervised the project. Z.W., S.W. and C.L. wrote the manuscript with input from all authors.

## Competing interests

C.L., Z.W., S.W., and H.K.-H.S. are named inventors on patent applications US2024/0202513A1 and CN118211616A, which cover aspects of the custom-designed analog spiking neuron circuits reported in this study. These patents are held by The University of Hong Kong. The remaining authors declare no competing interests.

Supplementary Information

# Fully Integrated Memristive Spiking Neural Network with Analog Neurons for High-Speed Event-Based Data Processing

Zhu Wang[1,3,#], Song Wang[1,#], Zhiyuan Du[1,3], Ruibin Mao[1,3], Yu Xiao[2], Hayden Kwok-Hay So[1], Peng Lin[2]*, and Can Li[1,3]*

[1]Department of Electrical and Electronic Engineering, The University of Hong Kong, Pokfulam, Hong Kong SAR, China
[2]College of Computer Science and Technology, Zhejiang University, Hangzhou, China
[3]Centre for Advanced Semiconductors and Integrated Circuits, The University of Hong Kong, Pokfulam, Hong Kong SAR, China
[#]These authors contributed equally

Email: penglin@zju.edu.cn; canl@hku.hk

## Contents

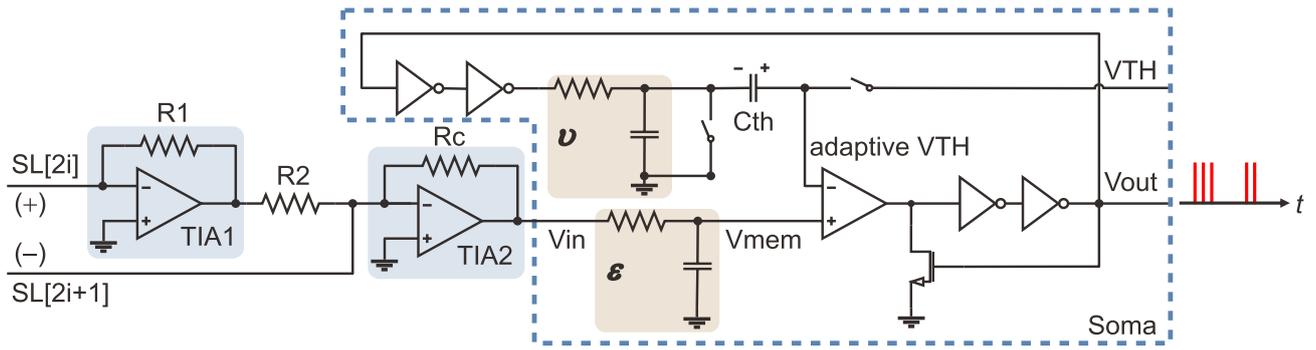

**Figure S1. Detailed schematics of the neuron circuit.** The design comprises a soma circuit and a differential TIA pair that converts incoming post-synaptic currents into a voltage. Resistors R1 and R2 are matched, while resistor Rc maps the post-synaptic voltage ("Vin") to the calculated post-synaptic vector-matrix multiplication result. The TIA2 confines Vin within the supply rails, thereby realizing a clipped activation function.

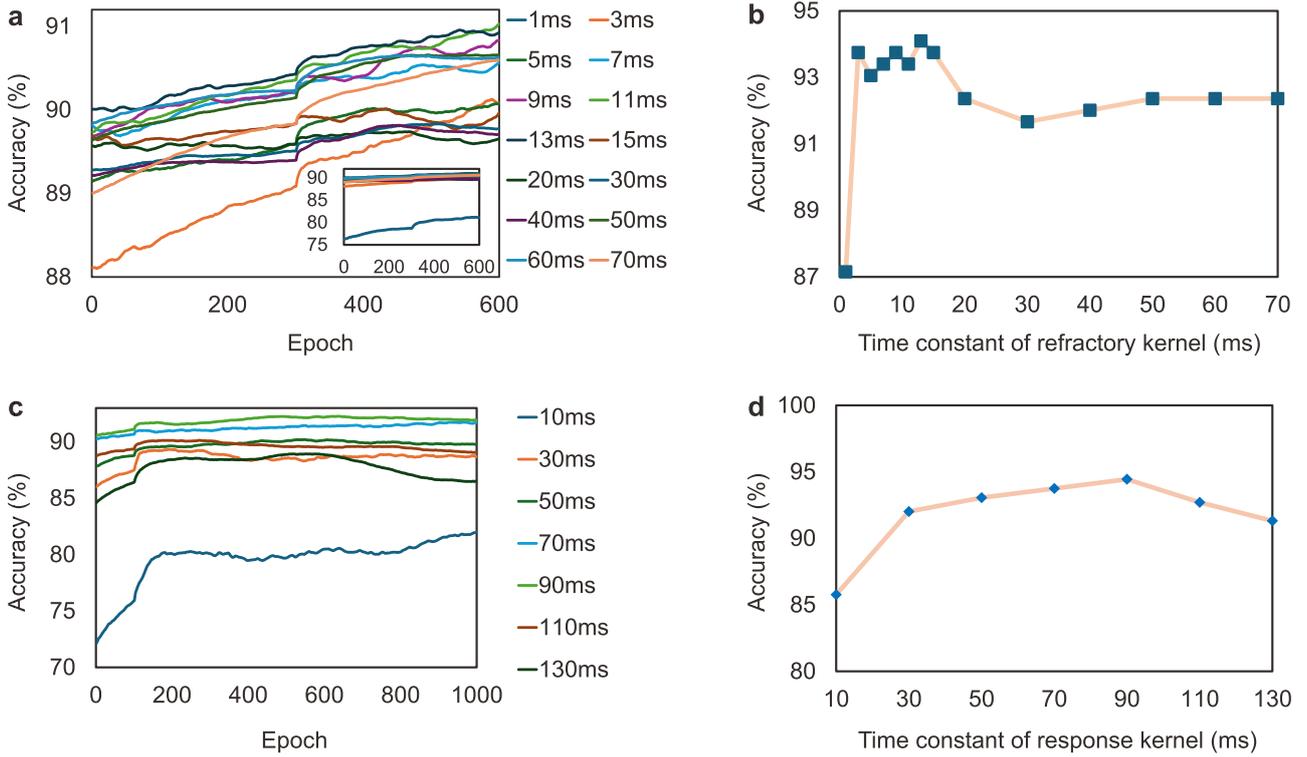

**Figure S2. Impact of neuron time constants on SNN system inference accuracy.** (a) Smoothed testing accuracy during training on the DVS128 Gesture dataset without data augmentation, varying the refractory kernel time constant ($\tau_r$) from 1 ms to 70 ms with a fixed response kernel time constant ($\tau_s$) of 50 ms. The inset shows a zoomed-out view of this plot. (b) Best testing accuracy from (a) versus $\tau_r$. A $\tau_r$ of 5 ms was selected to maintain a compact neuron circuit, yielding a $\tau_s:\tau_r$ ratio of 10:1. (c) Smoothed testing accuracy during training, varying $\tau_s$ from 10 ms to 130 ms while maintaining the $\tau_s:\tau_r$ ratio at 10:1. (d) Peak testing accuracy from (c) versus $\tau_s$. This panel, along with panel b, highlights the necessity of aligning neuronal time constants with the given task.

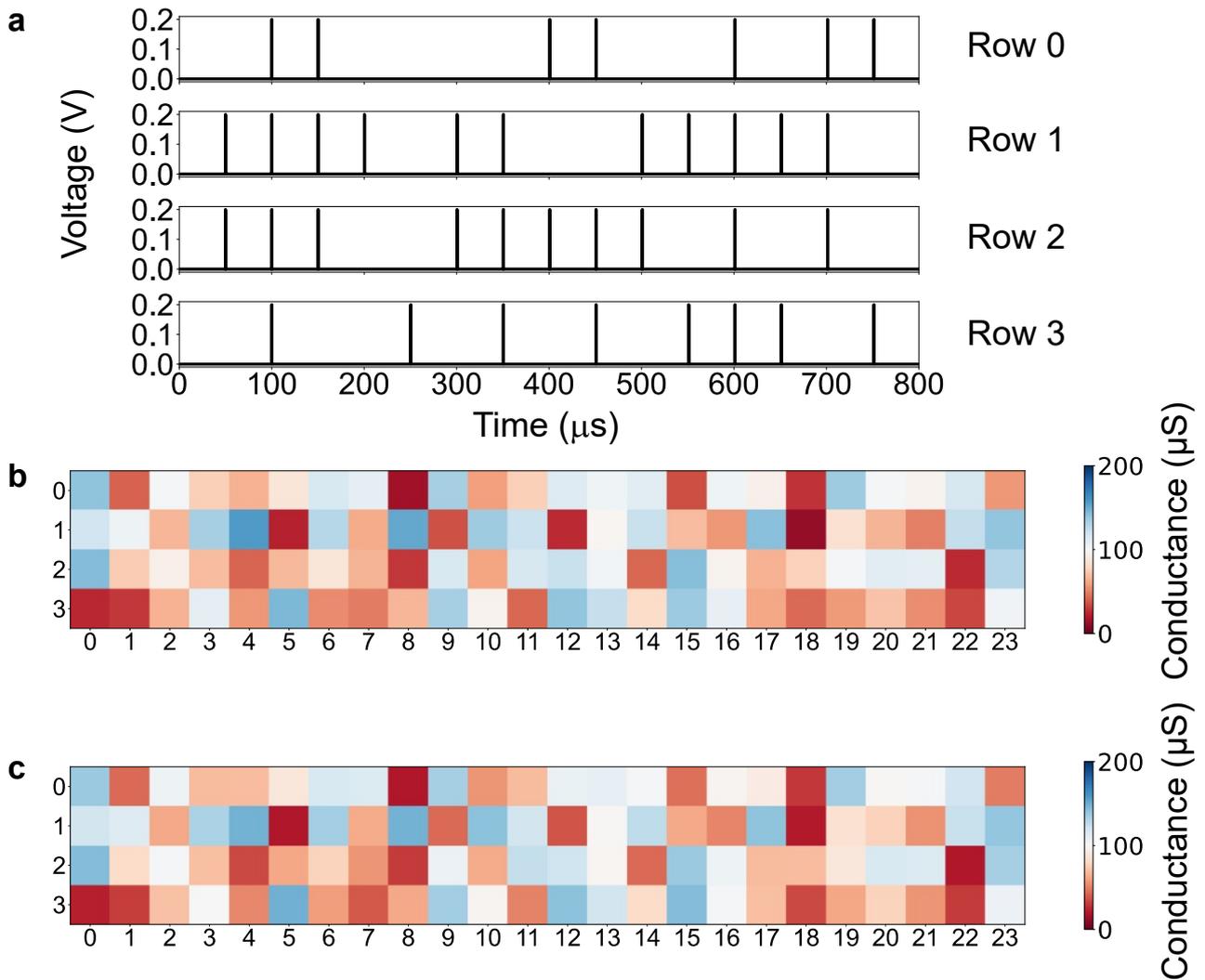

**Figure S3. (a) Input spike trains used in the synaptic computation experiment in Fig. 3(e, f), (b) Corresponding readout conductances of the programmed array, and (c) Target conductance values.** The post-synaptic voltage (Fig. 3f) was measured from column 0 of the array.

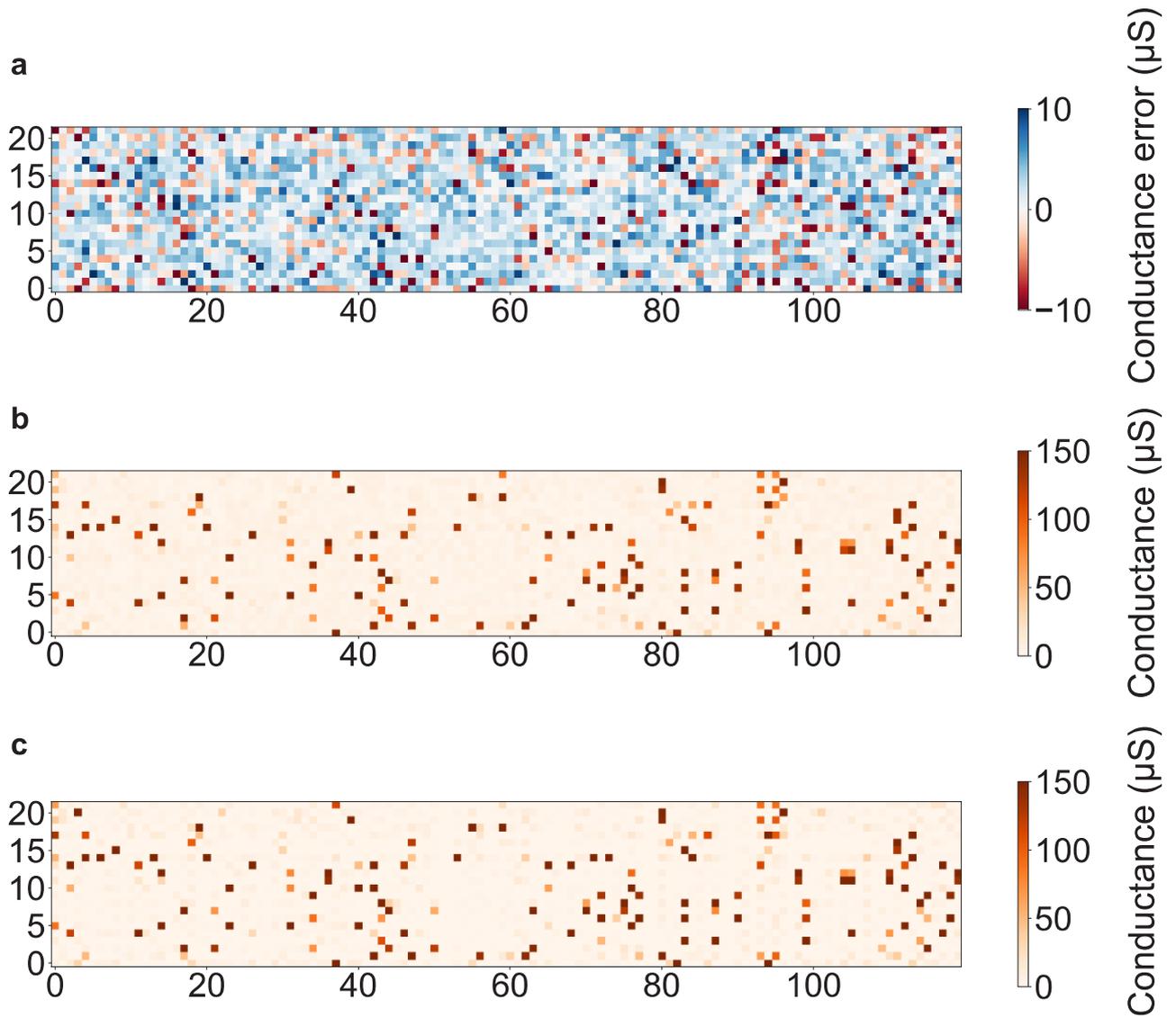

**Figure S4. Readout conductance in the final layer for DVS128 Gesture classification.** (a) Map of conductance errors across the 22×120 readout matrix. (b) The corresponding readout conductance values. (c) Target conductance values.

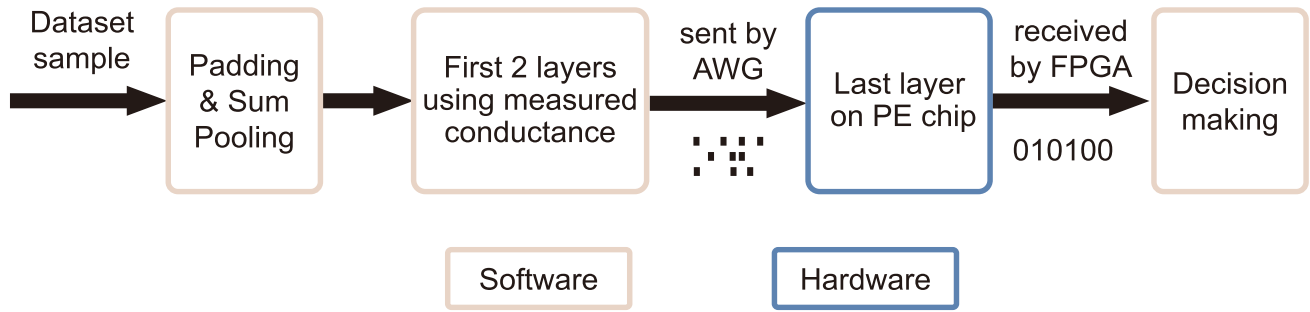

**Figure S5. Detailed implementation of the classification experiments on the DVS128 Gesture and NMNIST datasets.** The first two layers were performed in software using measured readout conductance, while the final layer was implemented with the PE macro chip.

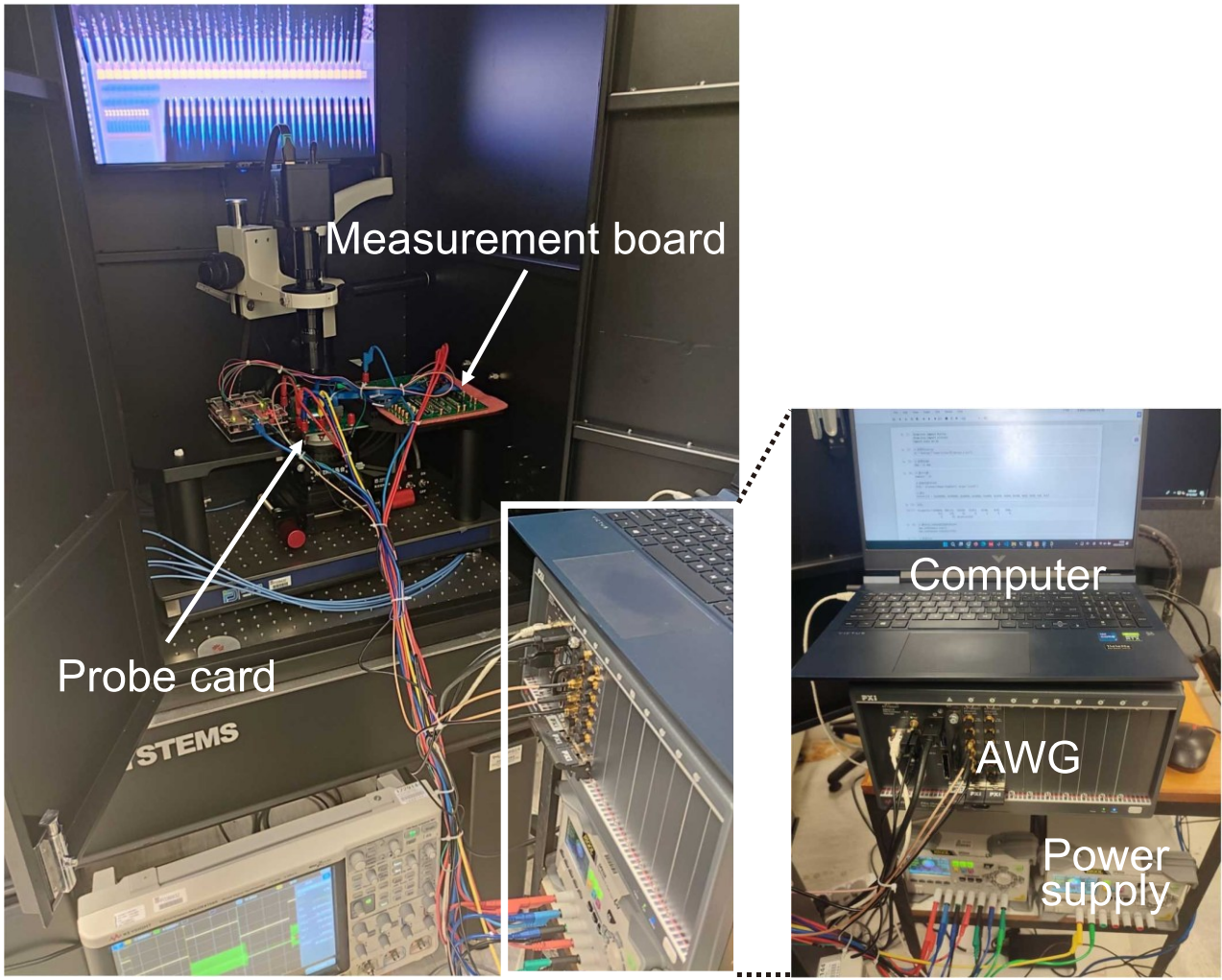

**Figure S6. Experimental setup for measuring the system inference accuracy.** The left image shows the experimental apparatus, including the probe station and measurement equipment. The inset shows the connected computer, AWG, and power supply.

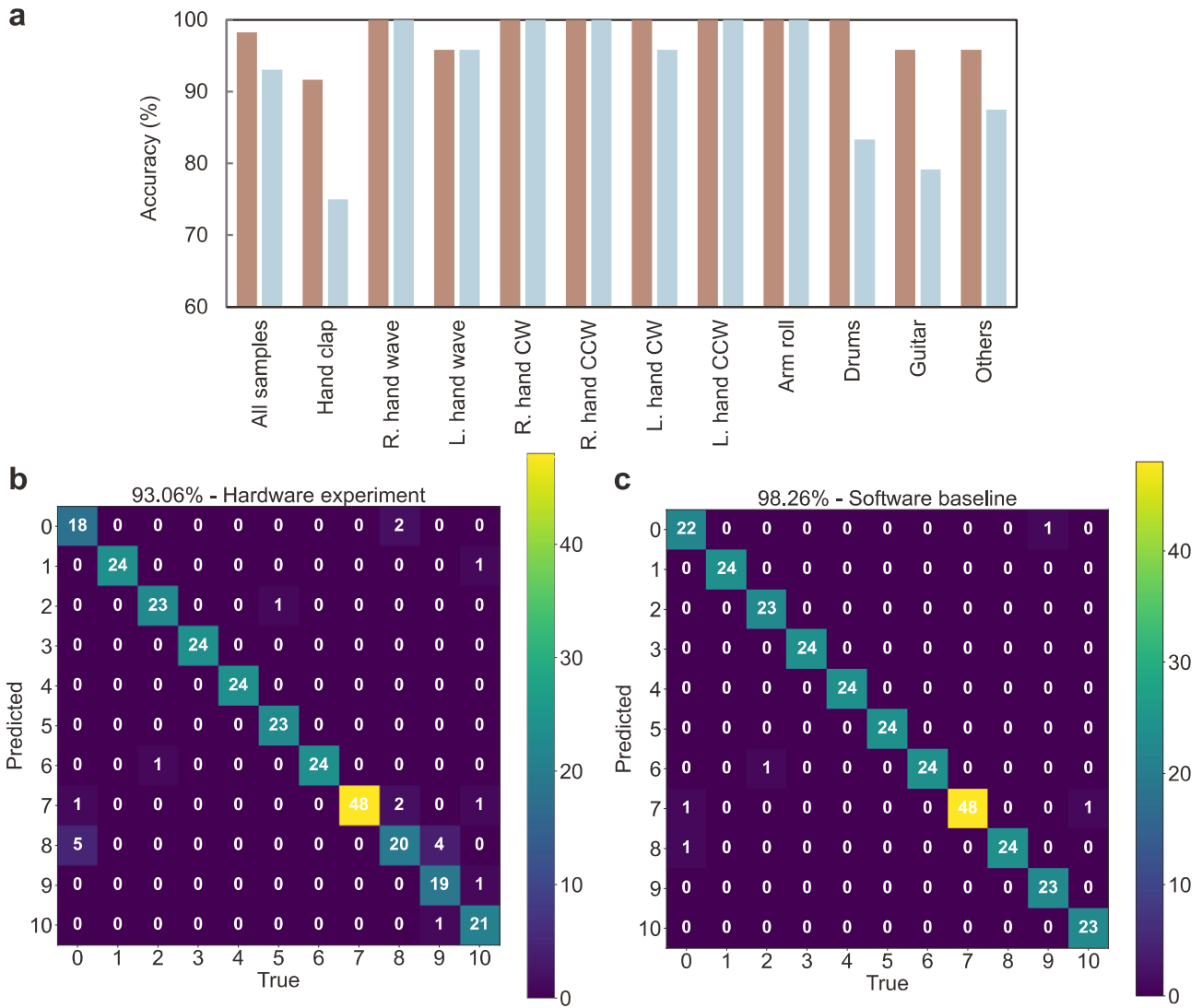

**Figure S7. Comparison of experimental classification accuracy to the software baseline on the DVS128 Gesture dataset.** (a) Experimental inference accuracy for each class (blue bars) versus the algorithm baseline (brown bars). The experimental accuracy (93.06%) decreases by 5.2% from the baseline (98.26%). (b) Confusion matrix for the experimental result, showing detailed classification, where gestures are labeled 0-10 (Hand clap, R. hand wave, L. hand wave, R. hand CW, R. hand CCW, L. hand CW, L. hand CCW, Arm roll, Drums, Guitar, Others). (c) Confusion matrix for the software baseline.

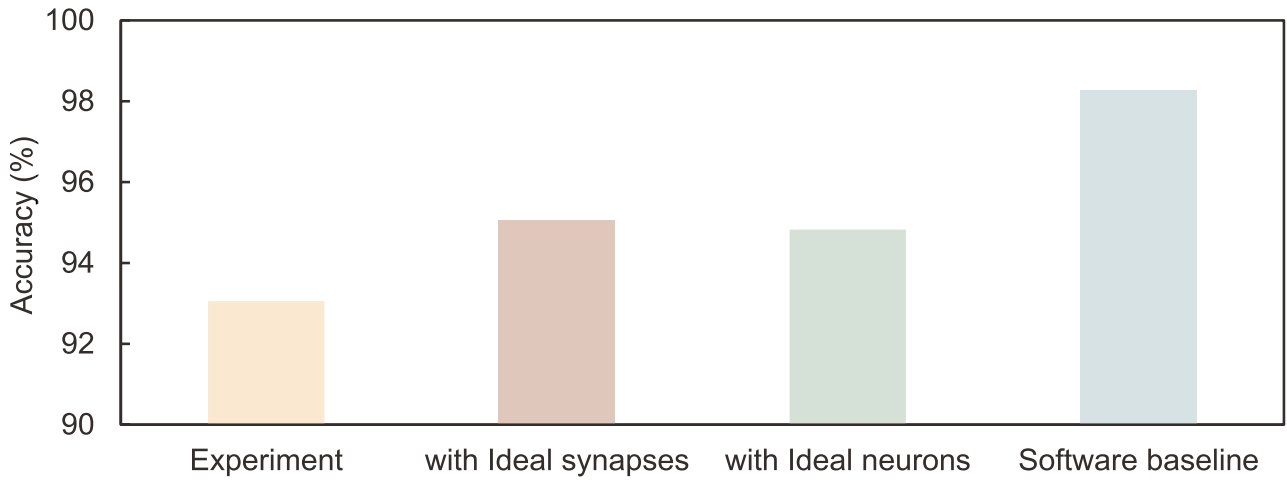

**Figure S8. RRAM synapse variation and neuron circuit non-idealities reduce system inference accuracy.** Benchmarked against the software baseline on the DVS128 Gesture dataset over ten inferences, a 3.19% average accuracy drop is caused by non-ideal neuron circuits ("with Ideal synapses"), while a 3.43% average accuracy drop is due to RRAM synapse variation ("with Ideal neurons"), compared with the experimental accuracy drop of 5.20% when both RRAM synapse variation and neuron circuit non-idealities are present.

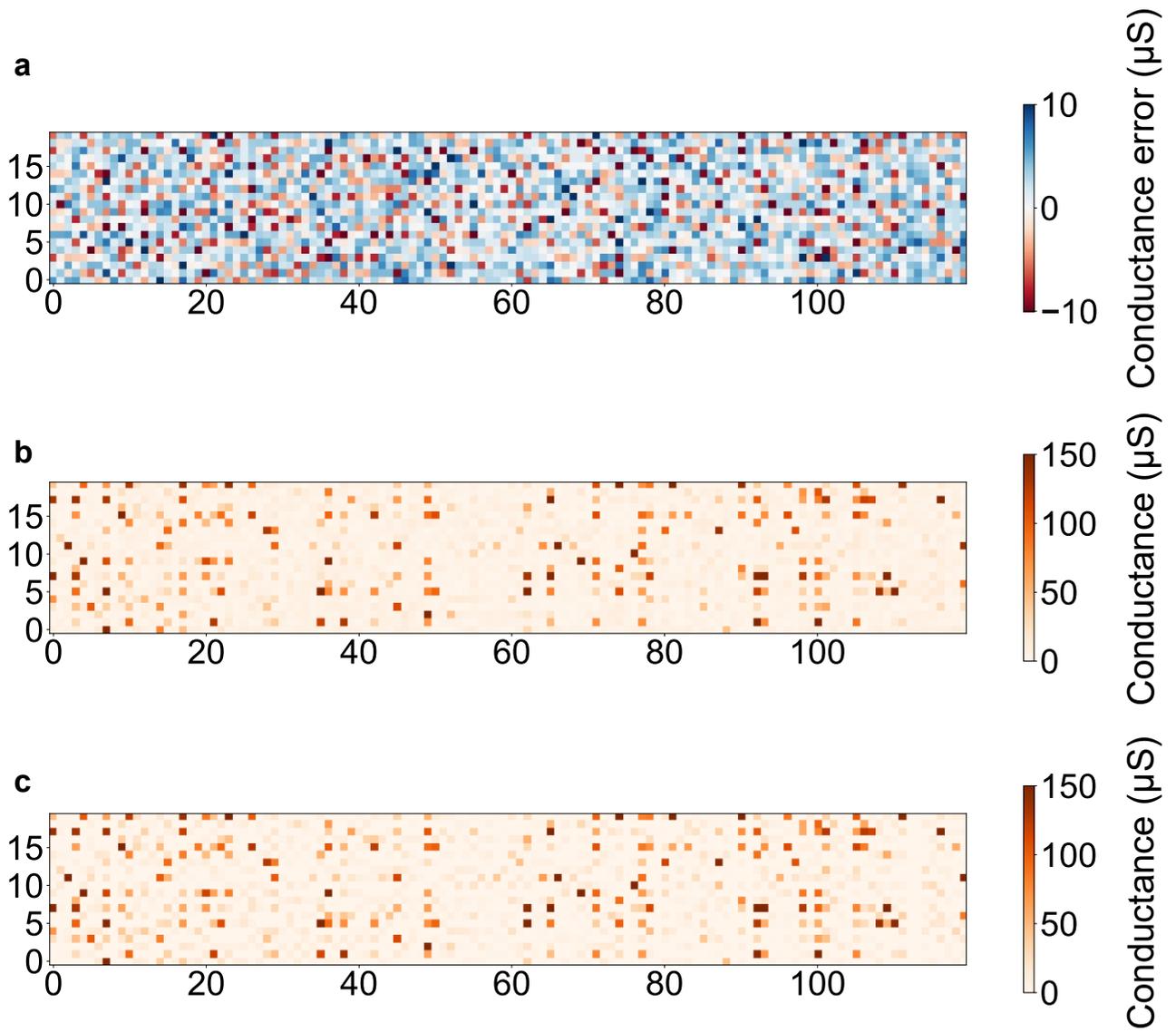

**Figure S9. Readout conductance in the final layer for NMNIST classification.** (a) Map of conductance errors across the 20×120 readout matrix. (b) The corresponding readout conductance values. (c) Target conductance values.

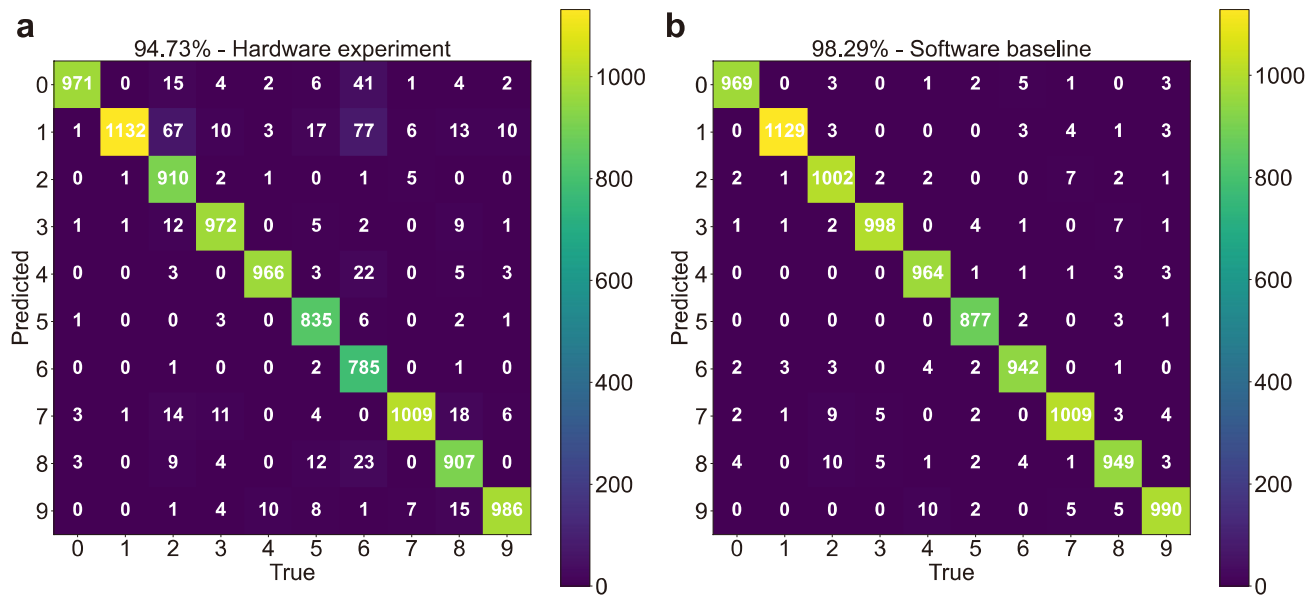

**Figure S10. Experimental accuracy compared to the software baseline for classifying digits on the NMNIST dataset.** (a) Confusion matrix for the experimental classification result. (b) Confusion matrix of the software baseline.

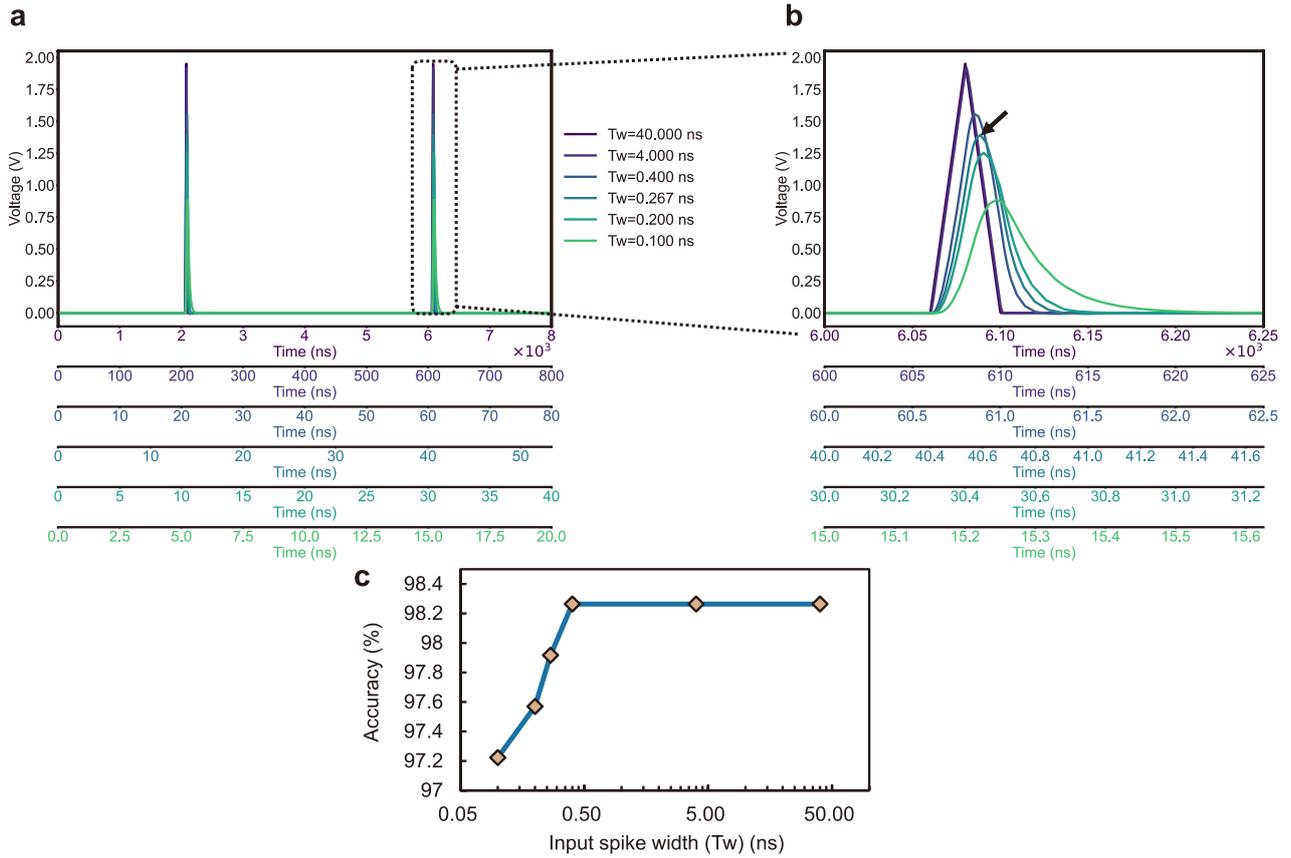

**Figure S11. Response of the RRAM crossbar array to input spikes of varying widths.** (a) Post-layout simulation results demonstrating the array's capability to process input spikes with widths ("Tw") down to 267 ps (arrow). (b) Zoomed-in view of the resultant output spike (Vin) attenuation with decreasing input spike width (Tw). (c) Impact of reduced input spike width (Tw) on the inference accuracy, decreasing from 98.26% to 97.92% at the narrowest supported spike width of 267 ps.

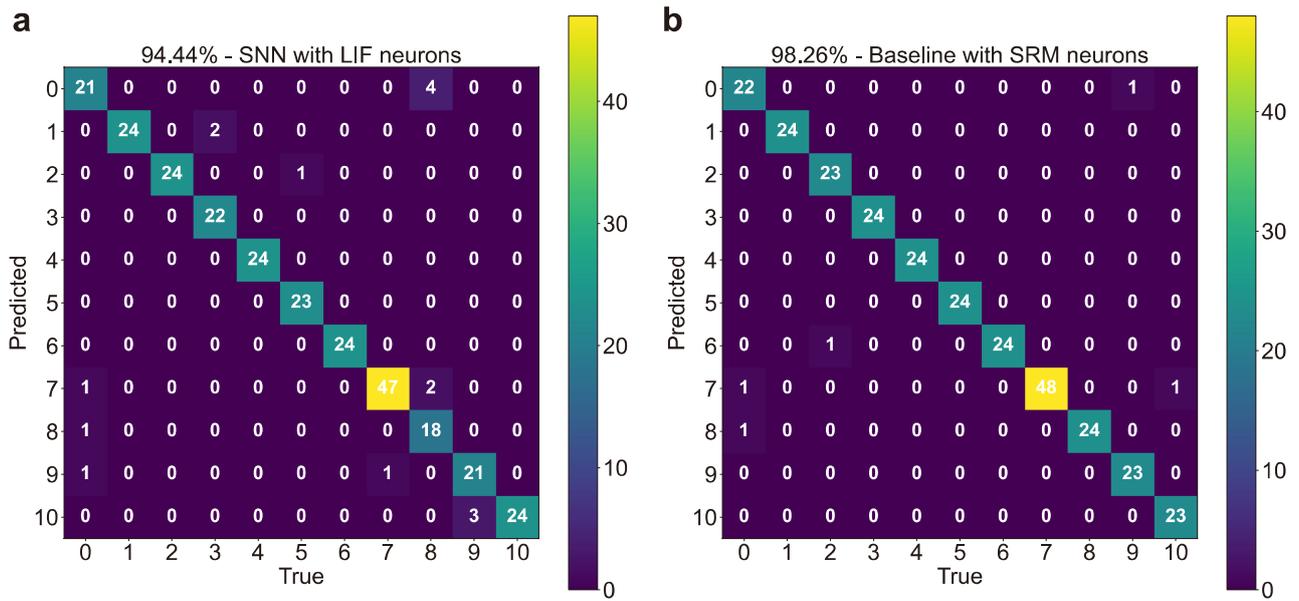

**Figure S12. Comparative analysis of algorithmic performance.** (a) Replacing SRM neurons with LIF neurons in an otherwise unchanged network structure results in a peak accuracy of 94.44%. (b) The baseline SNN with SRM neurons achieves a higher peak accuracy of 98.26%.

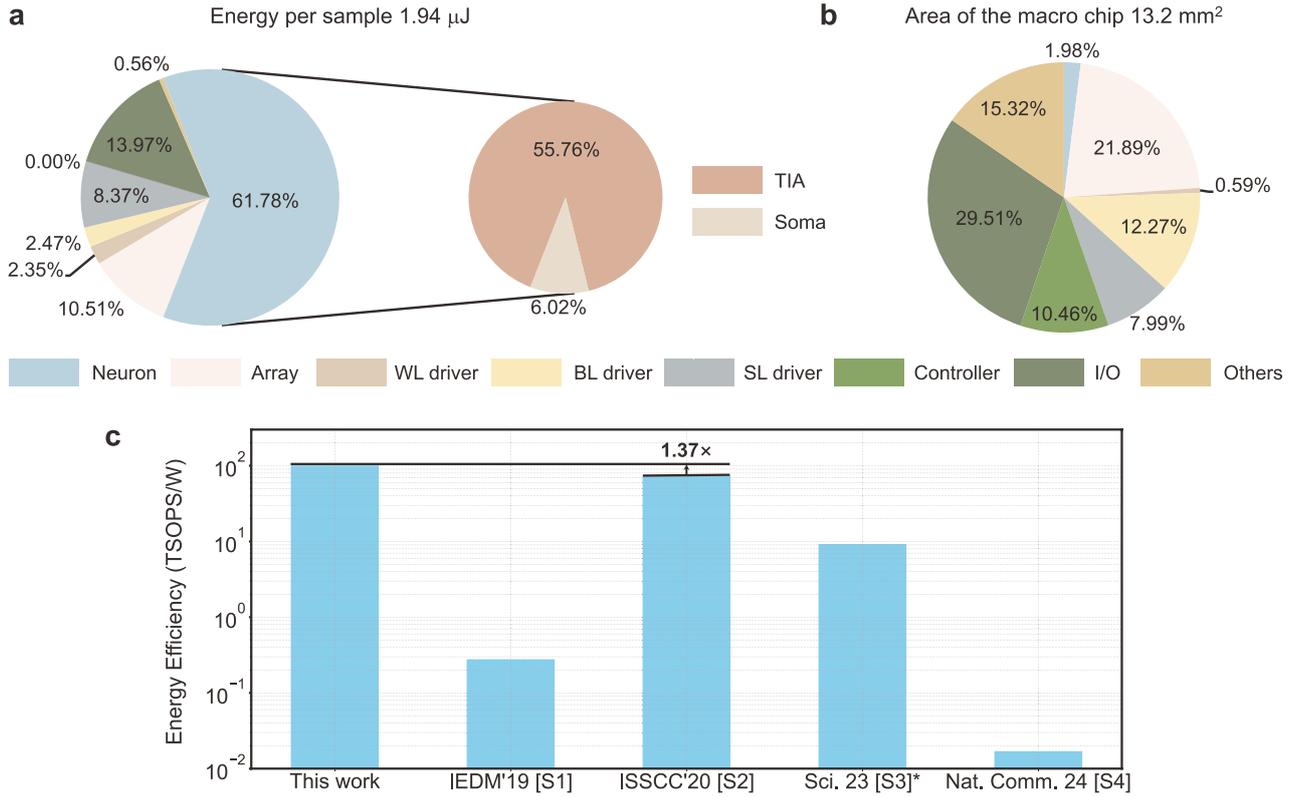

**Figure S13. Energy/area breakdown of the PE macro and energy-efficiency benchmarking.** (a) Energy breakdown during inference on the DVS128 Gesture dataset: TIAs within the neuron circuits dominate energy consumption (55.76%). (b) Area breakdown of the PE macro: neuron circuits occupy a relatively small footprint (1.98%) of the die, compared to the RRAM array (21.89%). (c) Energy-efficiency comparison with recent neuromorphic systems that integrate RRAM synapses with CMOS neurons [S1-4]. Our design achieves a 1.37× enhancement in energy efficiency compared to the prior SOTA. For consistency, one multiply-accumulate (MAC) counts as one synaptic operation, and each synaptic operation counts as two operations. * Neuron type: analog ReLU; included for the energy-efficiency benchmarking.

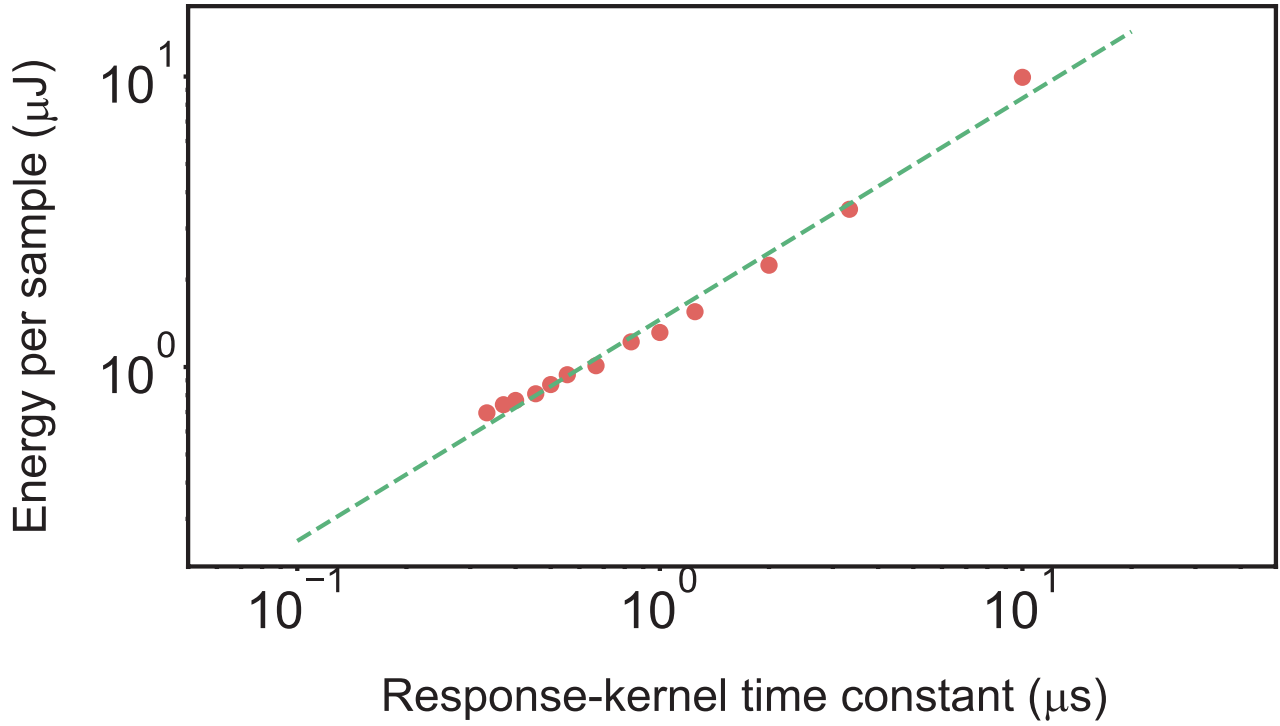

**Figure S14. Scalability analysis of the SNN system.** Energy consumed per inference sample in the PE macros falls as the response-kernel time constant ($\tau_s$) is further downscaled, while the ratio of response to refractory kernel time constants ($\tau_s:\tau_r$) is maintained at 10:1.

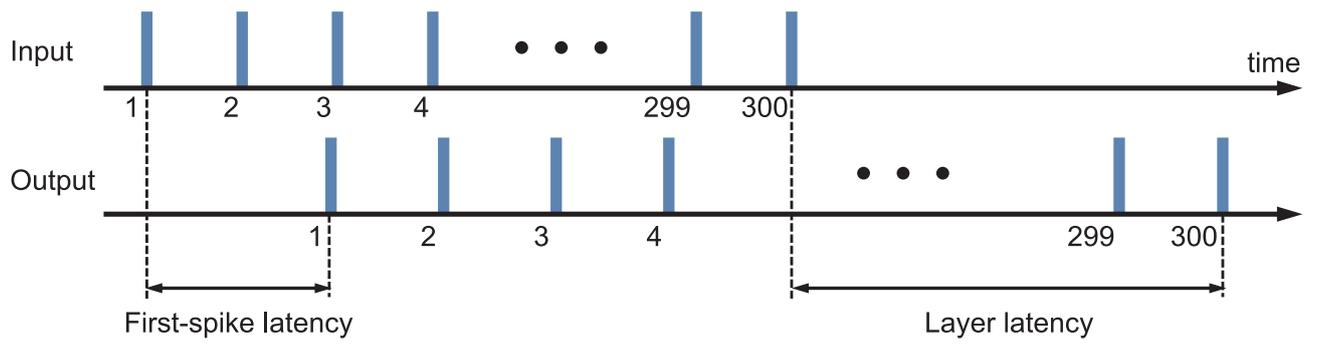

**Figure S15.** Schematic illustrating the latency incurred by a single layer within the SNN system.

**Supplementary Table S1. Benchmarking of algorithm performance on the DVS128 Gesture dataset.**

| Method | Network | Structure | Neuron | Parameters | Accuracy (%) |
|---|---|---|---|---|---|
| mMND (BPTT) [S5] | SNN | CONV | LIF | 1.1M | 98.0 |
| BPTT+PLIF [S6] | SNN | CONV | Parametric LIF | 6.7M | 97.57 |
| FPTT+LTC [S7] | SNN | CONV | Liquid Time-Constant Spiking Neuron | 6.7M | 97.22 |
| OTTT [S8] | SNN | CONV | LIF | 9.2M | 96.88 |
| mMND (STDP) [S5] | SNN | CONV | LIF | 0.81M | 96.6 |
| DECOLLE [S9] | SNN | CONV | LIF | 1.6M | 95.54±0.16 |
| DVSNet [S10] | SNN | CONV | LIF | 94K | 95.15 [a] |
| SLAYER [S11] | SNN | CONV | SRM | 1.1M | 93.64±0.49 |
| EGRU [S12] | RNN | GRU | – | 4.8M | 97.8 |
| LSTM [S13] | RNN | MLP | – | 4.2M | 86.81 |
| **This work** | SNN | MLP | SRM | 443K | 97.71±0.47 |

[a] Hardware-aware training

Our SNN model attained a peak accuracy of 98.26%, with an average accuracy of 97.71±0.47% over five trials.

**Supplementary Note 1. Rationale for employing backpropagation with surrogate gradients in SNN training.**

The training paradigm chosen for a SNN determines the network's achievable accuracy and latency. While several methods exist, our work adopts direct training with backpropagation through surrogate gradients. This choice is predicated on an analysis of the primary alternatives, biologically-inspired local learning and ANN-to-SNN conversion, whose respective limitations make them less suitable for achieving the SOTA performance required by our objectives in accuracy and latency.

Biologically-inspired pure Hebbian rules like spike-timing-dependent plasticity are efficient, highly valuable for on-chip learning. However, being fundamentally local and unsupervised, they lack a mechanism for global error correction. This locality makes them difficult to optimize the entire network for precise, complex tasks, resulting in a performance ceiling where their accuracy still lags behind supervised methods.

ANN-to-SNN conversion, another popular approach, leverages mature ANN training by translating a pre-trained ANN into an SNN [S14, 15]. While this can yield high accuracy, to accurately approximate the continuous activations of an ANN, the converted SNN has traditionally required long inference times (many time steps), which diminishes the latency and energy advantages inherent to spiking computation. Although recent work aims to reduce this latency, the paradigm is fundamentally designed to replicate ANN behavior through rate coding, rather than to natively exploit the rich temporal dynamics SNNs can offer. This makes conversion suboptimal for tasks where information is encoded in the precise timing of individual spikes.

To overcome these limitations, we employ backpropagation with surrogate gradients [S16-18]. This approach adapts the cornerstone of modern deep learning—global optimization via backpropagation—for direct SNN training. By doing so, it directly addresses the shortcomings of the other methods:

**(1) Overcomes local learning limits**

Unlike Hebbian rules, it provides global credit assignment that optimizes the entire network, closing the accuracy gap to ANN baselines.

**(2) Enables native temporal learning**

In contrast to ANN-to-SNN conversion, direct training allows the SNNs to learn features directly from temporal spike streams. This makes the SNNs effective to process event-based data and discover novel temporal patterns, fully leveraging the inherent computational strengths of SNNs [S6].

**(3) Achieves low-latency performance**

By training the SNNs end-to-end in the spiking domain, this method is not constrained by the inference time of conversion. It can achieve high accuracy with short inference time, leading to the low latency and high efficiency that SNNs promise.

While challenges remain in matching the performance of deep ANNs, direct backpropagation-based training represents an effective and direct path to training high-performance SNNs that are optimized for low-latency inference.

**Supplementary Note 2. On-chip integration for enhanced energy efficiency of PE macros.**

A detailed examination of the energy distribution within the PE macros (Fig. S13a) reveals that neuronal TIAs are the primary energy consumers, followed by the I/O elements. Currently, spike outputs are measured off-chip, and the associated parasitic limits how much neuronal time constants can be scaled down.

However, integrating spike capture on-chip can bypass these external limitations and enable more direct neuronal connections to the measurement. This integration offers the potential for a further reduction in neuronal time constants, thus enabling faster input stream processing and decreasing the per-sample energy consumption of PE macros (Fig. S14), although the overall energy per sample inference would depend on the efficiency of on-chip data processing and power management. Additionally, the area breakdown of our PE chip (Fig. S13b) underscores the same implication: neurons occupy a relatively small portion of the chip (1.98%), whereas I/O elements constitute the majority (29.51%)

## Supplementary References